%% file: main.tex
\documentclass{article}
\usepackage[utf8]{inputenc}

\usepackage[a-1b]{pdfx}   
\usepackage{graphicx}
\usepackage{subcaption}
\usepackage{times}
\usepackage{epsfig}
\usepackage{color}
\usepackage{xcolor}
\usepackage{graphicx}
\usepackage{wrapfig}
\usepackage{amssymb, amsmath}
\usepackage{multirow}
\usepackage{framed}
\usepackage{url}

\usepackage{listings}
\usepackage{booktabs} 
\usepackage{enumitem}
\usepackage{comment}
\usepackage{graphicx}
\usepackage[noend]{algpseudocode}
\usepackage{algorithm}
\usepackage{longtable}
\usepackage{tabularx}
\usepackage{listings}

\definecolor{codegreen}{rgb}{0,0.6,0}
\definecolor{codegray}{rgb}{0.5,0.5,0.5}
\definecolor{codepurple}{rgb}{0.58,0,0.82}
\definecolor{backcolour}{rgb}{0.95,0.95,0.92}
\definecolor{codekeyword}{RGB}{153, 51, 153}

\newcommand{\pythoninline}[1]{{\mintinline{python}{#1}}}

\usepackage[frozencache,cachedir=.]{minted}

\setminted[python]{
    bgcolor=backcolour,
    fontsize=\small,
    baselinestretch=1.15,
    linenos
}

\usepackage[framemethod=TikZ]{mdframed}
\mdfdefinestyle{style1}{
innerleftmargin=0.3cm,innerrightmargin=0.3cm,
innertopmargin=0.3cm ,innerbottommargin=0.3cm,
roundcorner=8pt,linewidth=0.6pt,
footnoteinside=false}

\usepackage{listings}
\lstdefinestyle{mystyle}{
  backgroundcolor=\color{backcolour},   
  commentstyle=\bfseries\color{codegreen},
  keywordstyle=\bfseries\color{codekeyword},
  numberstyle=\tiny\color{codegray},
  stringstyle=\color{codepurple},
  basicstyle=\ttfamily\footnotesize\linespread{1.15},
  breakatwhitespace=false,         
  breaklines=true,                 
  captionpos=b,                    
  keepspaces=true,                 
  numbers=left,                    
  numbersep=5pt,                  
  showspaces=false,                
  showstringspaces=false,
  showtabs=false,                  
  tabsize=2
}
\title{Automated Support for Unit Test Generation \\ A Tutorial Book Chapter}
\author{Afonso Fontes, Gregory Gay, Francisco Gomes de Oliveira Neto, Robert Feldt}
\date{From ``Optimising the Software Development Process with Artificial Intelligence'' \\(Springer, 2022)}

\begin{document}

\maketitle

\begin{abstract}

Unit testing is a stage of testing where the smallest segment of code that can be tested in isolation from the rest of the system---often a class---is tested. Unit tests are typically written as executable code, often in a format provided by a unit testing framework such as \texttt{pytest} for Python. 

Creating unit tests is a time and effort-intensive process with many repetitive, manual elements. To illustrate how AI can support unit testing, this chapter introduces the concept of \textit{search-based unit test generation}. This technique frames the selection of test input as an optimization problem---\textit{we seek a set of test cases that meet some measurable goal of a tester}---and unleashes powerful \textit{metaheuristic} search algorithms to identify the best possible test cases within a restricted timeframe. This chapter introduces two algorithms that can generate \texttt{pytest}-formatted unit tests, tuned towards coverage of source code statements. The chapter concludes by discussing more advanced concepts and gives pointers to further reading for how artificial intelligence can support  developers and testers when unit testing software.

\end{abstract}

\section{Introduction}\label{sec:intro}
\input{intro}

\section{Example System---BMI Calculator}\label{sec:bmi}
\input{sut_explanation}

\section{Unit Testing}\label{sec:unit_testing}
\input{unit_testing}

\section{Search-Based Test Generation}\label{sec:search_overview}
\input{search}

\section{Advanced Concepts}\label{sec:advanced}
\input{advanced}

\section{Conclusion}\label{sec:conclusion}
\input{conclusion}

\bibliographystyle{unsrt}
\bibliography{main}

\end{document}

%% file: intro.tex
Unit testing is a stage of testing where the smallest segment of code that can be tested in isolation from the rest of the system---often a class---is tested. Unit tests are typically written as executable code, often in a format provided by a unit testing framework such as \texttt{pytest} for Python. Unit testing is a popular practice as it enables test-driven development---where tests are written before the code for a class, and because the tests are often simple, fast to execute, and effective at verifying low-level system functionality. By being executable, they can also be re-run repeatedly as the source code is developed and extended.

However, creating unit tests is a time and effort-intensive process with many repetitive, manual elements. If elements of unit test creation could be automated, the effort and cost of testing could be significantly reduced. Effective automated test generation could also complement manually written test cases and help ensure test suite quality. Artificial intelligence (AI) techniques, including optimization, machine learning, natural language processing, and others, can be used to perform such automation.

To illustrate how AI can support unit testing, we introduce in this chapter the concept of \textit{search-based unit test input generation}. This technique frames the selection of test input as an optimization problem---\textit{we seek a set of test cases that meet some measurable goal of a tester}---and unleashes powerful \textit{metaheuristic} search algorithms to identify the best possible test input within a restricted timeframe. To be concrete, we use metaheuristic search to produce \texttt{pytest}-formatted unit tests for Python programs.

This chapter is laid out as follows:
\begin{itemize}
    \item In Section~\ref{sec:bmi}, we introduce our running example, a Body Mass Index (BMI) calculator written in Python.
    \item In Section~\ref{sec:unit_testing}, we give an overview of unit testing and test design principles. Even if you have prior experience with unit testing, this section provides an overview of the terminology we use.
    \item In Section~\ref{sec:search_overview}, we introduce and explain the elements of search-based test generation, including solution representation, fitness (scoring) functions, search algorithms, and the resulting test suites.
    \item In Section~\ref{sec:advanced}, we present advanced concepts that build on the foundation laid in this chapter.
\end{itemize}

To support our explanations, we created a Python project composed of (i) the class that we aim to test, (ii) a set of test cases created manually for that class following good practices in unit test design, and (iii) a simple framework including two search-based techniques that can generate new unit tests for the class. 

The code examples are written in Python 3, therefore, you must have Python 3 installed on your local machine in order to execute or extend the code examples. We target the \texttt{pytest} unit testing framework for Python.\footnote{For more information, see \url{https://pytest.org}.} We also make use of the \texttt{pytest-cov} plug-in for measuring code coverage of Python programs\footnote{See \url{https://pypi.org/project/pytest-cov/} for more information.}, as well as a small number of additional dependencies. All external dependencies that we rely on in this chapter can be installed using the \texttt{pip3} package installer included in the standard Python installation. Instructions on how to download and execute the code examples on your local machine are available in our code repository at \url{https://github.com/Greg4cr/PythonUnitTestGeneration}.

%% file: sut_explanation.tex
We illustrate concepts related to automated unit test generation using a class that implements different variants of a body mass index (BMI) classification\footnote{This is a relatively simple program compared to what is typically developed and tested in the software industry. However, it allows clear presentation of the core concepts of this chapter. After reading this chapter, you should be able to apply these concepts to a more complex testing reality.}. BMI is a value obtained from a person's weight and height to classify them as underweight, normal weight, overweight, or obese. There are two core parts to the BMI implementation: (i) the BMI value, and (ii) the BMI classification. The BMI value is calculated according to Equation~\ref{eq:bmi-value} using height in meters (m) and weight in kilos (kg). 

\begin{equation} \label{eq:bmi-value}
    BMI = \frac{weight}{(height)^2}
\end{equation}

The formula can be adapted to be used with different measurement systems (e.g., pounds and inches). In turn, the BMI classification uses the BMI value to classify individuals based on different threshold values that vary based on the person's age and gender\footnote{Threshold values can also vary depending on different continents or regions.}.

\begin{figure}
	\centering
\begin{minted}{python}
class BMICalc:
    def __init__(self, height, weight, age):
        self.height = height
        self.weight = weight
        self.age    = age

    def bmi_value(self):
        # The height is stored as an integer in cm. Here we convert it to
        # meters (m).
        bmi_value = self.weight / ((self.height / 100.0) ** 2)
        return bmi_value
    
    def classify_bmi_adults(self):
        if self.age > 19:
            bmi_value = self.bmi_value()
            if bmi_value < 18.5:
                return "Underweight"
            elif bmi_value < 25.0:
                return "Normal weight"
            elif bmi_value < 30.0:
                return "Overweight"
            elif bmi_value < 40.0:
                return "Obese"
            else:
                return "Severely Obese"
        else:
            raise ValueError(
                "Invalid age. The adult BMI classification requires an age "
                "older than 19.")
    \end{minted}
	\caption{An excerpt of the BMICalc class. The snippet includes the constructor for the BMICalc class, the method that calculates the BMI value according to Equation~\ref{eq:bmi-value}, and a method that returns the BMI classification for adults.}
	\label{fig:bmi_method}
\end{figure}

The BMI thresholds for children and teenagers vary across different age ranges (e.g., from 4 to 19 years old). As a result, the branching options quickly expand. In this example, we focus on the World Health Organization (WHO) BMI thresholds for cisgender\footnote{An individual whose personal identity and gender corresponds with their birth sex.} women, who are adults older than 19 years old\footnote{See \url{https://www.euro.who.int/en/health-topics/disease-prevention/nutrition/a-healthy-lifestyle/body-mass-index-bmi}}, and children\slash teenagers between 4 and 19 years old\footnote{See \url{https://www.who.int/tools/growth-reference-data-for-5to19-years/indicators/bmi-for-age}}. In Figure~\ref{fig:bmi_method}, we show an excerpt of the BMICalc class and the method that calculates the BMI value for adults. The complete code for the BMICalc class can be found at \url{https://github.com/Greg4cr/PythonUnitTestGeneration/blob/main/src/example/bmi_calculator.py}.

\begin{figure}
	\centering
\begin{minted}{python}
    def classify_bmi_teens_and_children(self):
        if self.age < 2 or self.age > 19:
            raise ValueError(
                'Invalid age. The children and teen BMI classification ' +
                'only works for ages between 2 and 19.')

        bmi_value = self.bmi_value()
        if   self.age <= 4:  ...
        elif self.age <= 7:  ...
        elif self.age <= 10: ...
        elif self.age <= 13: ...
        elif self.age <= 16: ...
        elif self.age <= 19: ...
    \end{minted}
	\caption{Method for the BMI classification of several age brackets that, in turn, expand further into the branches of each BMI classification. For readability, the actual thresholds were omitted from the excerpt above.}
	\label{fig:bmi_teens}
\end{figure}

The BMI classification is particularly interesting case for testing because, (i), it has \textit{numerous branching} statements based on multiple input arguments (age, height, weight, etc.), and (ii), it requires testers to think of specific \textit{combinations of all arguments} to yield BMI values able to cover all possible classifications. Table \ref{tab:bmi_values} shows all of the different thresholds for the BMI classification used in the BMICalc class.

\begin{table}[]
    \centering
    \caption{Threshold values used between different BMI classifications across the various age brackets. The children and teens reference values are for young girls.}
    \label{tab:bmi_values}
    \begin{tabularx}{\textwidth}{lXXXXXXX}
    \toprule
         \textbf{Classification} & $[2, 4]$ & $(4, 7]$ & $(7, 10]$ & $(10, 13]$ & $(13, 16]$ & $(16, 19]$ & $ > 19$ \\
         \midrule
Underweight    & $ \leq 14   $ & $ \leq 13.5 $ & $ \leq 14 $ & $ \leq 15   $ & $ \leq 16.5 $ & $ \leq 17.5 $ & $ < 18.5   $ \\
Normal weight  & $ \leq 17.5 $ & $ \leq 14   $ & $ \leq 20 $ & $ \leq 22   $ & $ \leq 24.5 $ & $ \leq 26.5 $ & $ < 25     $ \\
Overweight     & $ \leq 18.5 $ & $ \leq 20   $ & $ \leq 22 $ & $ \leq 26.5 $ & $ \leq 29   $ & $ \leq 31   $ & $ < 30     $ \\
Obese          & $ >    18.5 $ & $ >    20   $ & $ >    22 $ & $ >    26.5 $ & $ >    29   $ & $ >    31   $ & $ < 40     $ \\
Severely obese & ---           & ---           & ---         & ---           & ---           & ---           & $ \geq 40  $ \\
    \bottomrule
    \end{tabularx}
\end{table}

While the numerous branches add complexity to writing unit tests for our case example, the use of only integer input simplifies the problem. Modern software requires complex inputs of varying types (e.g., DOM files, arrays, abstract data types) which often need contextual knowledge from different domains such as automotive, web or cloud systems or embedded applications to create. In unit testing, the goal is to test small, isolated units of functionality that are often implemented as a collection of methods that receive primitive types as input. Next, we will discuss the scope of unit testing in detail, along with examples of good unit testing design practices, as applied to our BMI example.

%% file: unit_testing.tex
\begin{figure}[!t]
\centering
\includegraphics[width=0.5\textwidth]{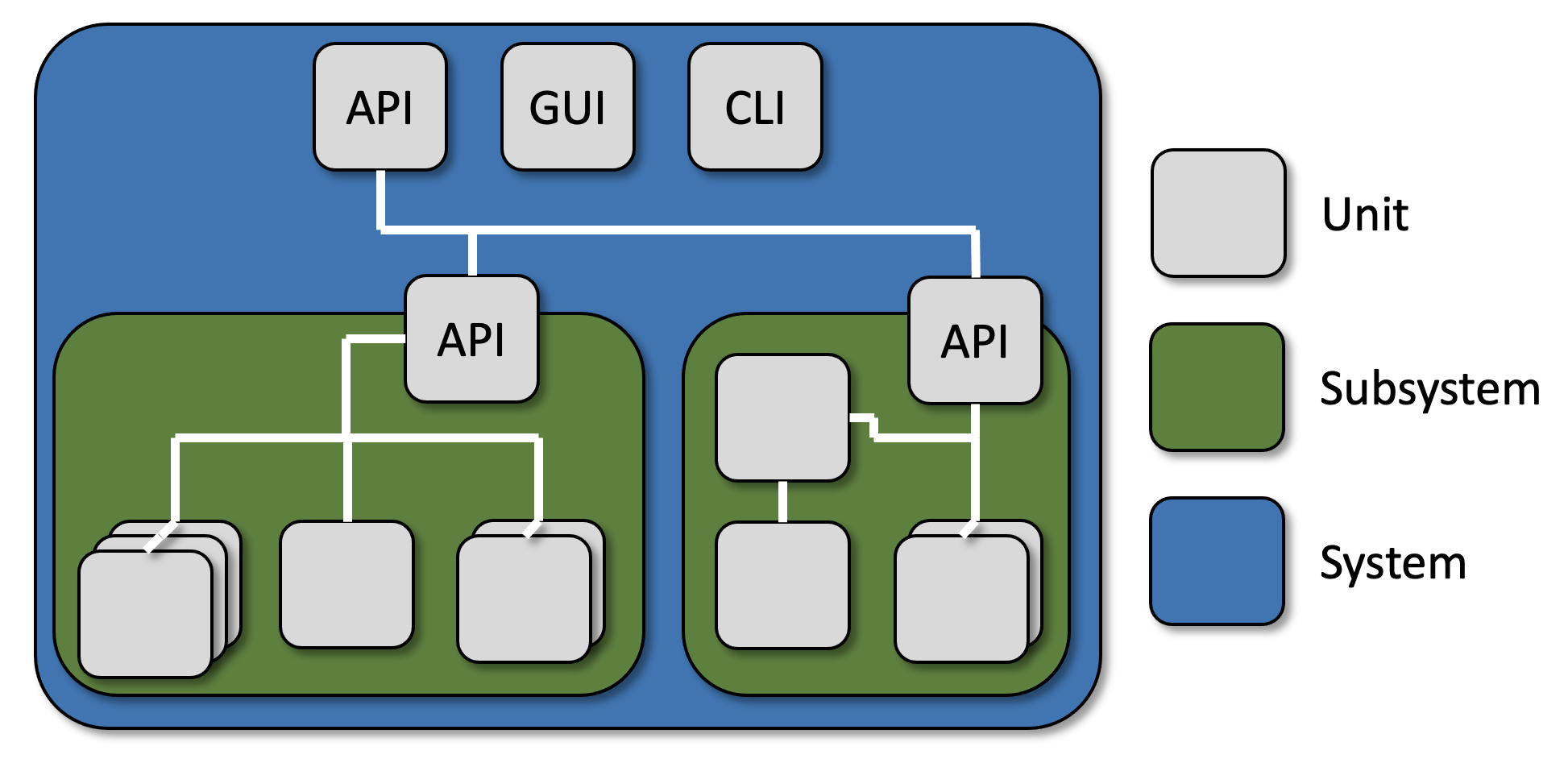}
\caption{Illustration of common levels of granularity in testing. A system is made up of one or more largely-independent subsystems. A subsystem is made up of one or more low-level ``units'' that can be tested in isolation.}
\label{fig:granularity}
\end{figure}

Testing can be performed at various levels of granularity, based on how we interact with the system-under-test (SUT) and the type of code structure we focus on. 
As illustrated in Figure~\ref{fig:granularity}, a system is often architected as a set of one or more cooperating or standalone subsystems, each responsible for a portion of the functionality of the overall system. Each subsystem, then, is made up of one or more ``units''---small, largely self-contained pieces of the system that contain a small portion of the overall system functionality. Generally, a unit is a single class when using object-oriented programming languages like Java and Python. 

Unit testing is the stage of testing where we focus on each of those individual units and test their functionality in \textit{isolation} from the rest of the system. The goal of this stage is to ensure that these low-level pieces of the system are trustworthy before they are integrated to produce more complex functionality in cooperation. If individual units seem to function correctly in isolation, then failures that emerge at higher levels of granularity are likely to be due to errors in their \textit{integration} rather than faults in the underlying units.  

Unit tests are typically written as executable code in the language of the unit-under-test (UUT). Unit testing frameworks exist for many programming languages, such as JUnit for Java, and are integrated into most development environments. Using the structures of the language and functionality offered by the unit testing framework, developers construct \textit{test suites}---collections of test cases---by writing test case code in special test classes within the source code. When the code of the UUT changes, developers can re-execute the test suite to make sure the code still works as expected. One can even write test cases before writing the unit code. Before the unit code is complete, the test cases will fail. Once the code is written, passing test cases can be seen as a sign of successful unit completion.

In our BMI example, the UUT is the \texttt{BMICalc} class outlined in the previous section. This example is written in Python. There are multiple unit testing frameworks for Python, with \texttt{pytest} being one of the most popular. We will focus on \texttt{pytest}-formatted test cases for both our manually-written examples and our automated generation example. Example test cases for the BMI example can be found at \url{https://github.com/Greg4cr/PythonUnitTestGeneration/blob/main/src/example/test_bmi_calculator_manual.py}, and will be explained below. 

Unit tests are typically the majority of tests written for a project. For example, Google recommends that approximately 70\% of test cases for Android projects be unit tests~\cite{Google20:AndroidTesting}. The exact percentage may vary, but this is a reasonable starting point for establishing your expectations. This split is partially, of course, due to the fact that there are more units than subsystem or system-level interfaces in a system and almost all classes of any importance will be targeted for unit testing. In addition, unit tests carry the following advantages:
\begin{itemize}
    \item \textbf{Useful Early in Development:} Unit testing can take place before development of a ``full'' version of a system is complete. A single class can typically be executed on its own, although a developer may need to \textit{mock} (fake the results of) its dependencies.
    \item \textbf{Simplicity:} The functionality of a single unit is typically more limited than a subsystem or the system as a whole. Unit tests often require less setup and the results require less interpretation than other levels of testing. Unit tests also often require little maintenance as the system as a whole evolves, as they are focused on small portions of the system. 
    \item \textbf{Execute Quickly:} Unit tests typically require few method calls and limited communication between elements of a system. They often can be executed on the developer's computer, even if the system as a whole runs on a specialised device (e.g., in mobile development, system-level tests must run on an emulator or mobile device, while unit tests can be executed directly on the local computer). As a result, unit tests can be executed quickly, and can be re-executed on a regular basis as the code evolves.
\end{itemize} 

When we design unit tests, we typically want to \textit{test all ``responsibilities'' associated with the unit}. We examine the functionality that the unit is expected to offer, and ensure that it works as expected. If our unit is a single class, each ``responsibility'' is typically a method call or a short series of method calls. Each broad outcome of performing that responsibility should be tested---e.g., alternative paths through the code that lead to different normal or exceptional outcomes. If a method sequence could be performed out-of-order, this should be attempted as well. We also want to \textit{examine how the ``state'' of class variables can influence the outcome of method calls}. Classes often have a set of variables where information can be stored. The values of those variables can be considered as the current state of the class. That state can often influence the outcome of calling a method. Tests should place the class in various states and ensure that the proper method outcome is achieved. 
When designing an individual unit test, there are typically five elements that must be covered in that test case:
\begin{itemize}
    \item \textbf{Initialization (Arrange):} This includes any steps that must be taken before the core body of the test case is executed. This typically includes initializing the UUT, setting its initial state, and performing any other actions needed to execute the tested functionality (e.g., logging into a system or setting up a database connection).
    \item \textbf{Test Input (Act):} The UUT must be forced to take actions through method calls or assignments to class variables. The test input consists of values provided to the parameters of those method calls or assignments. 
    \item \textbf{Test Oracle (Assert):} A test oracle, also known as an expected output, is used to validate the output of the called methods and the class variables against a set of encoded expectations in order to issue a verdict---pass or fail---on the test case. In a unit test, the oracle is typically formulated as a series of \textit{assertions} about method output and class attributes. An assertion is a Boolean predicate that acts as a check for correct behavior of the unit. The evaluation of the predicate determines the verdict (outcome) of the test case.
    \item \textbf{Tear Down (Cleanup):} Any steps that must be taken after executing the core body of the test case in order to prepare for the next test. This might include cleaning up temporary files, rolling back changes to a database, or logging out of a system.
    \item \textbf{Test Steps (Test Sequence, Procedure):} Code written to apply input to the methods, collect output, and compare the output to the expectations embedded in the oracle.
\end{itemize}

Unit tests are generally written as methods in dedicated classes grouping the unit tests for a particular UUT. The unit test classes are often grouped in a separate folder structure, mirroring the source code folder structure. For instance, the \texttt{utils.BMICalc} class stored in the \texttt{src} folder may be tested by a \texttt{utils.TestBMICalc} test class stored in the \texttt{tests} folder. The test methods are then executed by invoking the appropriate unit testing framework through the IDE or the command line (e.g., as called by a continuous integration framework). Figure~\ref{fig:testcase} shows four examples of test methods for the BMICalc class. Each test method checks a different scenario cover different aspects of good practices in unit test design, as will be detailed below. The test methods and scenarios are:
\begin{itemize}
    \item \pythoninline{test_bmi_value_valid()}: verifies the correct calculation of the BMI value for valid and typical ( ``normal'') inputs.
    \item \pythoninline{test_invalid_height()}: checks robustness for invalid values of height using exceptions.
    \item \pythoninline{test_bmi_adult()}: verifies the correct BMI classification for adults.
    \item \pythoninline{test_bmi_children_4y()}: checks the correct BMI classification for children up to 4 years old.
\end{itemize}



\begin{figure}[!t]
	\centering
\begin{minted}{python}
def test_bmi_value_valid():                          
    bmi_calc = bmi_calculator.BMICalc(150, 41, 18)   # Arrange
    bmi_value = bmi_calc.bmi_value()                 # Act
    # Here, the approx method allows 0.01 
    #  differences in floating point errors.
    assert pytest.approx(bmi_value, abs=0.1) == 18.2 # Assert.

# Cases expected to throw exception
def test_invalid_height():
    # 'with' blocks expect exceptions to be thrown, hence
    #   the assertion is checked *after* the constructor call 
    with pytest.raises(ValueError) as context:           # Assert
        bmi_calc = bmi_calculator.BMICalc(-150, 41, 18)  # Act

    with pytest.raises(ValueError) as context:           # Assert       
        bmi_calc.height = 0                              # Act

def test_bmi_adult():
    bmi_calc = bmi_calculator.BMICalc(160, 65, 21)  # Arrange
    bmi_class = bmi_calc.classify_bmi_adults()       # Act
    assert bmi_class == "Overweight"                # Assert

def test_bmi_children_4y():
    bmi_calc = bmi_calculator.BMICalc(100, 13, 4)
    bmi_class = bmi_calc.classify_bmi_teens_and_children()
    assert bmi_class == "Underweight"
    \end{minted}
	\caption{Examples of test methods for the BMICalc class using the pytest framework.}
	\label{fig:testcase}
\end{figure}

Due to the challenges in representing real numbers in binary computing systems, a good practice in unit test design is to allow for an error range when assessing the correct calculations of floating point arithmetic. We use the \pythoninline{approx} method from the pytest framework to automatically verify whether the returned value lies within the $0.1$ range of our test oracle. For instance, our first test case would pass if the returned BMI value would be $18.22$ or $18.25$, however, it would fail for $18.3$. Most unit testing frameworks provide a method to assert floating points within specific ranges. Testers should be careful when asserting results from floating point arithmetic because failures in those assertions can represent precision or range limitations in the programming language instead of faults in the source code, such as incorrect calculations. For instance, neglecting to check for float precision is a ``test smell'' that can lead to flaky test executions~\cite{Luo14:EAF,eck2019understanding}.\footnote{Tests are considered flaky if their verdict (pass or fail) changes when no code changes are made. In other words, the tests seems to show random behaviour.}
If care is not taken some tests might fail when running them on a different computer or when the operating system has been updated.

In addition to asserting the valid behaviour of the UUT (also referred informally to as ``happy paths''), unit tests should check the robustness of the implementation. For example, testers should examine how the class handles exceptional behaviour. There are different ways to design unit tests to handle exceptional behaviour, each with its trade-offs. One example is to use exception handling blocks and include \textit{failing assertions} (e.g., \pythoninline{assert false}) in points past the code that triggers an exception. However, those methods are not effective in checking whether specific types of exceptions have been thrown, such as distinguishing between input\slash output exceptions for ``file not found'' or database connection errors versus exceptions thrown due to division by zero or accessing null variables. Those different types of exceptions represent distinct types of error handling situations that testers may choose to cover in their test suites. Therefore, many unit test frameworks have methods to assert whether the UUT raises specific types of exception. Here we use the \pythoninline{pytest.raises(...)} context manager to capture the exceptions thrown when trying to specify invalid values for height and check whether they are the exceptions that we expected, or whether there are unexpected exceptions. Additionally, testers can include assertions to verify whether the exception includes an expected message.

One of the challenges in writing good unit tests is deciding on the maximum size and scope of a single test case. For instance, in our BMICalc class, the \pythoninline{classifyBMI_teensAndChildren()} method has numerous branches to handle the various BMI thresholds for different age ranges. Creating a single test method that exercises all branches for all age ranges would lead to a very long test method with dozens of assertions. This test case would be hard to read and understand. Moreover, such a test case would hinder debugging efforts  because the tester would need to narrow down which specific assertion detected a fault. Therefore, in order to keep our test methods small, we recommend breaking down test coverage of the method (\pythoninline{classifyBMI_teensAndChildren()}) into a series of small test cases---with each test covering a different age range. In turn, for improved coverage, each of those test cases should assert all BMI classifications for the corresponding age bracket.

Testers should avoid creating redundant test cases in order to improve the cost-effectiveness of the unit testing process. Redundant tests exercise the same behaviour, and do not bring any value (e.g., increased coverage) to the test suite. For instance, checking invalid height values in the \pythoninline{test_bmi_adult()} test case would introduce redundancy because those cases are already covered by the \pythoninline{test_invalid_height()} test case. On the other hand, the (\pythoninline{test_bmi_adult()}) test case currently does not attempt to invoke BMI for ages below 19. Therefore, we can improve our unit tests by adding this invocation to the existing test case, or---even better---creating a new method with that invocation (e.g., \pythoninline{test_bmi_adult_invalid()}).

\subsection{Supporting Unit Testing with AI}

Conducting rigorous unit testing can be an expensive, effort-intensive task. The effort required to create a single unit test may be negligible over the full life of a project, but this effort adds up as the number of classes increases. If one wants to test thoroughly, they may end up creating hundreds to thousands of tests for a large-scale project. Selecting effective test input and creating detailed assertions for each of those test cases is not a trivial task either. The problem is not simply one of scale. Even if developers and testers have a lot of knowledge and good intentions, they might forget or not have the time needed to think of all important cases. They may also cover some cases more than others, e.g., they might focus on valid inputs, but miss important invalid or boundary cases. The effort spent by developers does not end with test creation. Maintaining test cases as the SUT evolves and deciding how to allocate test execution resources effectively---deciding \textit{which} tests to execute---also require care, attention, and time from human testers. 

Ultimately, developers often make compromises if they want to release their product on time and under a reasonable budget. This can be problematic, as insufficient testing can lead to critical failures in the field after the product is released. Automation has a critical role in controlling this cost, and ensuring that both sufficient quality and quantity of testing is achieved. AI techniques---including optimization, machine learning, natural language processing, and other approaches---can be used to partially automate and support aspects of unit test creation, maintenance, and execution. For example,
\begin{itemize}
    \item Optimization and reinforcement learning can \textit{select test input} suited to meeting measurable testing goals. This can be used to create either new test cases or to amplify the effectiveness of human-created test cases.
    \item The use of supervised and semi-supervised machine learning approaches has been investigated in order to \textit{infer test oracles} from labeled executions of a system for use in judging the correctness of new system executions.
    \item Three different families of techniques, powered by optimization, supervised learning, and clustering techniques, are used to make effective use of computing resources when executing test cases:
    \begin{itemize}
        \item \textit{Test suite minimization} techniques suggest redundant test suites that could be removed or ignored during test execution. 
        \item \textit{Test case prioritization} techniques order test cases such that the potential for early fault detection or code coverage is maximised. 
        \item \textit{Test case selection} techniques identify the subset of test cases that relate in some way to recent changes to the code, ignoring test cases with little connection to the changes being tested.
    \end{itemize}
\end{itemize}

If aspects of unit testing---such as test creation or selection of a subset for execution---can be even partially automated, the benefit to developers could be immense. AI has been used to support these, and other, aspects of unit testing. In the remainder of this chapter, we will focus on \textbf{test input generation}. In Section~\ref{sec:advanced}, we will also provide pointers to other areas of unit testing that can be partially automated using AI.

Exhaustively applying all possible input is infeasible due to an enormous number of possibilities for most real-world programs and units we need to test. Therefore, deciding \textit{which} inputs to try becomes an important decision. Test generation techniques can create partial unit tests covering the initialization, input, and tear down stages. The developer can then supply a test oracle or simply execute the generated tests and capture any crashes that occur or exceptions that are thrown.  One of the more effective methods of automatically selecting effective test input is \textit{search-based test generation}. We will explain this approach in the following sections.

A word of caution, before we continue---it is our firm stance that AI \textit{cannot} replace human testers. The points above showcase a set of good practices for unit test design. Some of these practices may be more easily achieved by either a human or an intelligent algorithm. For instance, properties such as readability mainly depends on human comprehension. Choosing readable names or defining the ideal size and scope for test cases may be infeasible or difficult to achieve via automation. On the other hand, choosing inputs (values or method calls) that mitigate redundancy can be easily achieved through automation through instrumentation, e.g., the use of code coverage tools. 

AI can make unit testing more cost-effective and productive when used to support human efforts. However, there are trade-offs involved when deciding how much to rely on AI versus the potential effort savings involved. AI cannot replace human effort and creativity. However, it can reduce human effort on repetitive tasks, and can focus human testers towards elements of unit testing where their creativity can have the most impact. And over time, as AI-based methods become better and stronger, there is likely to be more areas of unit testing they can support or automate.

%% file: search.tex
Test input selection can naturally be seen as a search problem. When you create test cases, you often have one or more goals. Perhaps that goal is to find violations of a specification, to assess performance, to look for security vulnerabilities, to detect excessive battery usage, to achieve code coverage, or any number of other things that we may have in mind when we design test cases. We cannot try all input---any real-world piece of software with value has a near-infinite number of possible inputs we could try. However, somewhere in that space of possibilities lies a subset of inputs that best meets the goals we have in mind. Out of all of the test cases that could be generated for a UUT, we want to identify---systematically and at a reasonable cost---those that best meet those goals. Search-based test generation is an intuitive AI technique for locating those test cases that maps to the same process we might use ourselves to find a solution to a problem. 

Let us consider a situation where you are asked a question. If you do not know the answer, you might make a guess---either be an educated guess or one made completely at random. In either case, you would then get some feedback. \textit{How close were you to reaching the ``correct'' answer?} If your answer was not correct, you could then make a second guess. Your second guess, if nothing else, should be \textit{closer} to being correct based on the knowledge gained from the feedback on that initial guess. If you are still not correct, you might then make a third, fourth, etc. guess---each time incorporating feedback on the previous guess. 

Test input generation can be mapped to the same process. We start with a problem we want to solve. We have some goal that we want to achieve through the creation of unit tests. \textit{If that goal can be measured}, then we can automate input generation. Fortunately, many testing goals can be measured. 
\begin{itemize}
    \item If we are interested in exploring the exceptions that the UUT can throw, then we want the inputs that \textit{trigger the most exceptions}. 
    \item If we are interested in covering all outcomes of a function, then we can divide the output into representative values and identify the inputs that \textit{cover all representative output values}. 
    \item If we are interested in executing all lines of code, then we are searching for the inputs that \textit{cover more of the code structure}.
    \item If we are interested in executing a wide variety of input, then we want to find a set of inputs with \textit{the highest diversity in their values}. 
\end{itemize}
Attainment of many goals can be measured, whether as a percentage of a known checklist or just a count that we want to maximize. Even if we have a higher-level goal in mind that cannot be directly measured, there may be measurable sub-goals that correlate with that higher-level goal. For example, ``find faults'' cannot be measured---we do not know what faults are in our goal---but maximizing code coverage or covering diverse outputs may increase the likelihood of detecting a fault.

Once we have a measurable \textbf{goal}, we can automate the guess-and-check process outlined above via a metaheuristic optimization algorithm. \textbf{Metaheuristics} are strategies to sample and evaluate values during our search. Given a measurable goal, a metaheuristic optimization algorithm can systematically sample the space of possible test input, guided by feedback from one or more \textbf{fitness functions}---numeric scoring functions that judge the optimality of the chosen input based on its attainment of our goals. The exact process taken to sample test inputs from that space varies from one metaheuristic to another. However, the core process can be generically described as:

\begin{enumerate}
    \item Generate one or more initial \textbf{solutions} (test suites containing one or more unit tests).
    \item While time remains:
    \begin{enumerate}
        \item Evaluate each solution using the \textbf{fitness functions}.
        \item Use feedback from the fitness functions and the sampling strategy employed by the \textbf{metaheuristic} to improve the solutions. 
    \end{enumerate}
    \item Return the best solution seen during this process. 
\end{enumerate}

In other words, we have an optimization problem. We make a guess, get feedback, and then use that additional knowledge to make a \textit{smarter} guess. We keep going until we run out of time, then we work with the best solution we found during that process. 

The choice of both metaheuristic and fitness functions is crucial to successfully deploying search-based test generation. Given the existence of a near-infinite space of possible input choices, the order that solutions are tried from that space is the key to efficiently finding a solution. The metaheuristic---guided by feedback from the fitness functions---overcomes the shortcomings of a purely random input selection process by using a deliberate strategy to sample from the input space, gravitating towards ``good'' input and discarding input sharing properties with previously-seen ``bad'' solutions. By determining how solutions are evolved and selected over time, the choice of metaheuristic impacts the quality and efficiency of the search process. Metaheuristics are often inspired by natural phenomena, such as swarm behavior or evolution within an ecosystem. 

In search-based test generation, the fitness functions represent our goals and guide the search. They are responsible for evaluating the quality of a solution and offering feedback on how to improve the proposed solutions. Through this guidance, the fitness functions shape the resulting solutions and have a major impact on the quality of those solutions. Functions must be efficient to execute, as they will be calculated thousands of times over a search. Yet, they also must provide enough detail to differentiate candidate solutions and guide the selection of optimal candidates.

Search-based test generation is a powerful approach because it is \textit{scalable} and \textit{flexible}. Metaheuristic search---by strategically sampling from the input space---can scale to larger problems than many other generation algorithms. Even if the ``best'' solution can not be found within the time limit, search-based approaches typically can return a ``good enough'' solution. Many goals can be mapped to fitness functions, and search-based approaches have been applied to a wide variety of testing goals and scenarios. Search-based generation often can even achieve higher goal attainment than developer-created tests.

In the following sections, we will explain the highlighted concepts in more detail and explore how they can be applied to generate partial unit tests for Python programs. In Section~\ref{sec:solutions}, we will explain how to represent solutions. Then, in Section~\ref{sec:fitness}, we will explore how to represent two common goals as fitness functions. In Section~\ref{sec:algorithms}, we will explain how to use the solution representation and fitness functions as part of two common metaheuristic algorithms. Finally, in Section~\ref{sec:results}, we will illustrate the application of this process on our BMI example.

\subsection{Solution Representation}\label{sec:solutions}

When solving any problem, we first must define the form the solution to the problem must take. What, exactly, does a solution to a problem ``look'' like? What are its contents? How can it be manipulated? Answering these questions is crucial before we can define how to identify the ``best'' solution.

In this case, we are interested in identifying a set of unit tests that maximise attainment of a testing goal. This means that a solution is a \textbf{test suite}---a collection of test cases. We can start from this decision, and break it down into the composite elements relevant to our problem.
\begin{itemize}
    \item A solution is a \textbf{test suite}. 
    \item A test suite contains one or more \textbf{test cases}, expressed as individual methods of a single test class.
    \item The solution interacts with a \textbf{unit-under-test (UUT)} which is a single, identified Python class with a constructor (optional) and one or more methods.
    \item Each test case contains an \textbf{initialization} of the UUT which is a call to its constructor, if it has one.
    \item Each test case then contains one or more \textbf{actions}, i.e., calls to one of the methods of the UUT or assignments to a class variable.
    \item The initialization and each action have zero or more \textbf{parameters} (input) supplied to that action.
\end{itemize}

This means that we can think of a test suite as a collection of test cases, and each test case as a single initialization and a collection of actions, with associated parameters. When we generate a solution, we choose a number of test cases to create. For each of those test cases, we choose a number of actions to generate. Different solutions can differ in size---they can have differing numbers of test cases---and each test case can differ in size---each can contain a differing number of actions.

\begin{figure}
	\centering
	\begin{subfigure}{.45\textwidth}
      \begin{verbatim}
[
    [
        [-1, [246, 680, 2]], 
        [2, [18]], 
        [4, []], 
        [1, [466]], 
        [5, []], 
        [4, []], 
        [1, [26]], 
        [5, []]
    ]
]
    \end{verbatim}
\end{subfigure}%
\begin{subfigure}{.5\textwidth}
\begin{minted}{python}
import pytest
import bmi_calculator 

def test_0():
    cut = bmi_calculator.BMICalc(246,680,2)
    cut.age = 18
    cut.classify_bmi_teens_and_children()
    cut.weight = 466
    cut.classify_bmi_adults()
    cut.classify_bmi_teens_and_children()
    cut.weight = 26
    cut.classify_bmi_adults()
            
\end{minted}
\end{subfigure}%
	\caption{The genotype (internal, left) and phenotype (external, right) representations of a solution containing a single test case. Each identifier in the genotype is mapped to a function with a corresponding list of parameters. For instance, \texttt{1} maps to setting the \texttt{weight}, and \texttt{5} maps to calling the method \texttt{classify\_bmi\_adults()}}.
	\label{fig:representation}
\end{figure}

In search-based test generation, we represent two solutions in two different forms:
\begin{itemize}
    \item \textbf{Phenotype (External) Representation:} The phenotype is the version of the solution that will be presented to an external audience. This is typically in a human-readable form, or a form needed for further processing. 
    \item \textbf{Genotype (Internal) Representation:} The genotype is a representation used internally, within the metaheuristic algorithm. This version includes the properties of the solution that are relevant to the search algorithm, e.g., the elements that can be manipulated directly. It is generally a minimal representation that can be easily manipulated by a program.
\end{itemize}

Figure~\ref{fig:representation} illustrates the two representations of a solution that we have employed for unit test generation in Python. The phenotype representation takes the form of an executable pytest test class. In turn, each test case is a method containing an initialization, followed by a series of method calls or assignments to class variables. This solution contains a single test case, \pythoninline{test_0()}. It begins with a call to the constructor of the UUT, \pythoninline{BMICalc}, supplying a height of $246$, a weight of $680$, and an age of $2$. It then applies a series of actions on the UUT: setting the age to $18$, getting a BMI classification from \pythoninline{classify_bmi_teens_and_children()}, setting the weight to $466$, getting further classifications from each method, setting the weight to $26$, then getting one last classification from \pythoninline{classify_bmi_adults()}.

This is our desired external representation because it can be executed at will by a human tester, and it is in a format that a tester can read. However, this representation is not ideal for use by the metaheuristic search algorithm as it cannot be easily manipulated. If we wanted to change one method call to another, we would have to identify which methods were being called. If we wanted to change the value assigned to a variable, we would have to identify (a) which variable was being assigned a value, (b) identify the portion of the line that represents the value, and (c), change that value to another. Internally, we require a representation that can be manipulated quickly and easily. 

This is where the genotype representation is required. In this representation, a test suite is a \pythoninline{list} of test cases. If we want to add a test case, we can simply append it to the list. If we want to access or delete an existing test case, we can simply select an index from the list. Each test case is a \pythoninline{list} of actions. Similarly, we can simply refer to the index of an action of interest. 

Within this representation, each action is a \pythoninline{list} containing (a) an action identifier, and (b), a \pythoninline{list} of parameters to that action (or an empty \pythoninline{list} if there are no parameters). The action identifier is linked to a separate list of actions that the tester supplies, that stores the method or variable name and type of action, i.e., assignment or method call (we will discuss this further in Section~\ref{sec:results}). An identifier of $-1$ is reserved for the constructor. 

The solution illustrated in Figure~\ref{fig:representation} is not a particularly effective one. It consists of a single test case that applies seemingly random values to the class variables (the initial constructor creates what may be the world's largest two-year old). This solution only covers a small set of BMI classifications, and only a tiny portion of the branching behavior of the UUT. However, one could imagine this as a starting solution that could be manipulated over time into a set of highly effective test cases. By making adjustments to the genotype representation, guided by the score from a fitness function, we can introduce those improvements.

\subsection{Fitness Function}\label{sec:fitness}

\begin{figure}
	\centering
\begin{minted}{python}
def calculate_fitness(metadata, fitness_function, num_tests_penalty, 
                      length_test_penalty, solution):
    fitness = 0.0

    # Get the statement coverage over the code
    fitness += statement_fitness(metadata, solution)

    # Add a penalty to control test suite size
    fitness -= float(len(solution.test_suite) / num_tests_penalty)

    # Add a penalty to control the length of individual test cases
    # Get the average test suite length)
    total_length = 0
    total_length = sum([len(test) for test in solution.test_suite]) / len(
                       solution.test_suite)
    fitness -= float(total_length / length_test_penalty)

    solution.fitness = fitness
    \end{minted}
	\caption{The high-level calculation of the fitness function.}
	\label{fig:score}
\end{figure}

As previously-mentioned, fitness functions are the cornerstone of search-based test generation. The core concept is simple and flexible---a fitness function is simply a function that takes in a solution candidate and returns a ``score'' describing the quality of that solution. This gives us the means to differentiate one solution from another, and more importantly, to tell if one solution is \textit{better} than another. 

Fitness functions are meant to embody the goals of the tester. They tell us how close a test suite came to meeting those goals. The fitness functions employed determine what properties the final solution produced by the algorithm will have, and shape the evolution of those solutions by providing a target for optimization.

Essentially any function can serve as a fitness function, as long as it returns a numeric score. It is common to use a function that emits either a percentage (e.g., percentage of a checklist completed) or a raw number as a score, then either maximise or minimise that score. 
\begin{itemize}
    \item A fitness function should \textit{not} return a Boolean value. This offers almost no feedback to improve the solution, and the desired outcome may not be located.
    \item A fitness function should yield (largely) continuous scores. A small change in a solution should not cause a large change (either positive or negative) in the resulting score. Continuity in the scoring offers clearer feedback to the metaheuristic algorithm.
    \item The best fitness functions offer not just an indication of quality, but a \textit{distance} to the optimal quality. For example, rather than measuring completion of a checklist of items, we might offer some indication of how close a solution came to completing the remaining items on that checklist. In this chapter, we use a simple fitness function to clearly illustrate search-based test generation, but in Section~\ref{sec:advanced}, we will introduce a distance-based version of that fitness function.
\end{itemize}

Depending on the algorithm employed, either a single fitness function or multiple fitness functions can be optimised at once. We focus on single-function optimization in this chapter, but in Section~\ref{sec:advanced}, we will also briefly explain how multi-objective optimization is achieved. 

To introduce the concept of a fitness function, we utilise a fitness function based on the \textbf{code coverage} attained by the test suite. When testing, developers must judge:
(a) whether the produced tests are effective and (b) when they can stop writing additional tests. Coverage criteria provides developers with guidance on both of those elements. As we cannot know what faults exist without verification, and as testing cannot---except in simple cases---conclusively 
prove the absence of faults, these criteria are intended to serve as an approximation of efficacy. If the goals of the chosen criterion are met, then we have put in a measurable testing effort and can decide whether we have tested enough.

There are many coverage criteria, with varying levels of tool support. The most common criteria measure coverage of structural elements of the software, such as individual statements, branches of the software's control flow, and complex Boolean conditional statements. One of the most common, and most intuitive, coverage criteria is \textbf{statement coverage}. It simply measure the percentage of executable lines of code that have been triggered at least once by a test suite. The more of the code we have triggered, the more thorough our testing efforts are---and, ideally, the likely we will be to discover a fault. The use of statement coverage as a fitness function encourages the metaheuristic to explore the structure of the source code, reaching deeply into branching elements of that code.

As we are already generating \texttt{pytest}-compatible test suites, measuring statement coverage is simple. The pytest plugin \pythoninline{pytest-cov} measures statement coverage, as well as branch coverage---a measurement of how many branching control points in the UUT (e.g., \texttt{if}-statement and loop outcomes) have been executed---as part of executing a \texttt{pytest} test class. By making use of this plug-in, statement coverage of a solution can be measured as follows:

\begin{enumerate}
    \item Write the phenotype representation of the test suite to a file.
    \item Execute pytest, with the \texttt{--cov=<python file to measure coverage over>} command.
    \item Parse the output of this execution, extracting the percentage of coverage attained.
    \item Return that value as the fitness.
\end{enumerate}


This measurement yields a value between 0--100, indicating the percentage of statements executed by the solution. We seek to \textit{maximise} the statement coverage. Therefore, we employ the following formulation to obtain the fitness value of a test suite (shown as code in Figure~\ref{fig:score}):

\begin{equation} \label{eq:fitness}
    fitness(solution) = statement\_coverage(solution) - bloat\_penalty(solution)
\end{equation}

\noindent The \textit{bloat penalty} is a small penalty to the score intended to control the size of the produced solution in two dimensions: the number of test methods, and the number of actions in each test. A massive test suite may attain high code coverage or yield many different outcomes, but it is likely to contain many redundant elements as well. In addition, it will be more difficult to understand when read by a human. In particular, long sequences of actions may hinder efforts to debug the code and identify a fault. Therefore, we use the bloat penalty to encourage the metaheuristic algorithm to produce \textit{small-but-effective} test suites. The bloat penalty is calculated as follows:

\begin{equation} \label{eq:bloat}
\begin{aligned}
    bloat\_penalty(solution) & = (num\_test\_cases / num\_tests\_penalty) \\ 
    & + (average\_test\_length / length\_test\_penalty) \\
\end{aligned} 
\end{equation}

\noindent Where $num\_tests\_penalty$ is 10 and $length\_test\_penalty$ is 30. That is, we divide the number of test cases by 10 and the average length of a single test case (number of actions) by 30. These weights could be adjusted, depending on the severity of the penalty that the tester wishes to apply. It is important to not penalise too heavily, as that will increase the difficulty of the core optimization task---some expansion in the number of tests or length of a test is needed to cover the branching structure of the code. These penalty values allow some exploration while still encouraging the metaheuristic to locate smaller solutions.


\subsection{Metaheuristic Algorithms}\label{sec:algorithms}

Given a solution representation and a fitness function to measure the quality of solutions, the next step is to design an algorithm capable of producing the best possible solution within the available resources. Any UUT with reasonable complexity has a near-infinite number of possible test inputs that could be applied. We cannot reasonable try them all. Therefore, the role of the metaheuristic is to intelligently sample from that space of possible inputs in order to locate the best solution possible within a strict time limit. 

There are many metaheuristic algorithms, each making use of different mechanisms to sample from that space. In this chapter, we present two algorithms:

\begin{itemize}
    \item \textbf{Hill Climber:} A simple algorithm that produces a random initial solution, then attempts to find better solutions by making small changes to that solution---restarting if no better solution can be found.
    \item \textbf{Genetic Algorithm:} A more complex algorithm that models how populations of solutions evolve over time through the introduction of mutations and through the breeding of good solutions.
\end{itemize}

The Hill Climber is simple, fast, and easy to understand. However, its effectiveness depends strongly on the quality of the initial guess made. We introduce it first to explain core concepts that are built upon by the Genetic Algorithm, which is slower but potentially more robust. 

\subsubsection{Common Elements}

Before introducing either algorithm in detail, we will begin by discussing three elements shared by both algorithms---a metadata file that defines the actions available for the UUT, random test generation, and the search budget. 

\begin{figure}
\centering
\small
\begin{minted}{python}
{
    "file": "bmi_calculator",
    "location": "example/",
    "class": "BMICalc",
    "constructor": { 
        "parameters": [
            { "type": "integer", "min": -1 },
            { "type": "integer", "min": -1 },
            { "type": "integer", "min": -1, "max": 150 }
        ] },
    "actions": [
        { "name": "height", "type": "assign", "parameters": [ 
            { "type": "integer", "min": -1 } ] 
        },
        { "name": "weight", "type": "assign", "parameters": [ 
            { "type": "integer", "min": -1 } ] 
        },
        { "name": "age", "type": "assign", "parameters": [ 
            { "type": "integer", "min": -1, "max": 150 } ] 
        },
        { "name": "bmi_value", "type": "method" },
        { "name": "classify_bmi_teens_and_children", "type": "method" },
        { "name": "classify_bmi_adults", "type": "method" }
    ]
}
\end{minted}
\caption{Metadata definition for class \pythoninline{BMICalc}.}
\label{fig:metadata}
\end{figure}

\smallskip\noindent\textbf{UUT Metadata File:} To generate unit tests, the metaheuristic needs to know \textit{how} to interact with the UUT. In particular, it needs to know what methods and class variables are available to interact with, and what the parameters of the methods and constructor are. To provide this information, we define a simple JSON-formatted metadata file. The metadata file for the BMI example is shown in Figure~\ref{fig:metadata}, and we define the fields of the file as follows:
\begin{itemize}
    \item \textbf{file:} The python file containing the UUT.
    \item \textbf{location:} The path of the file.
    \item \textbf{class:} The name of the UUT.
    \item \textbf{constructor:} Contains information on the parameters of the constructor.
    \item \textbf{actions:} Contains information about each action.
    \begin{itemize}
        \item \textbf{name:} The name of the action (method or variable name).
        \item \textbf{type:} The type of action (\pythoninline{method} or \pythoninline{assign}).
        \item \textbf{parameters:} Information about each parameter of the action.
        \begin{itemize}
            \item \textbf{type:} Datatype of the parameter. For this example, we only support integer input. However, the example code could be expanded to handle additional datatypes.
            \item \textbf{min:} An optional minimum value for the parameter. Used to constrain inputs to a defined range.
            \item \textbf{max:} An optional maximum value for the parameter. Used to constrain inputs to a defined range.
        \end{itemize}
    \end{itemize}
\end{itemize}

This file not only tells the metaheuristic what actions are available for the UUT, it suggests a starting point for ``how'' to test the UUT by allowing the user to optionally constrain the range of values.
This allows more effective test generation by limiting the range of guesses that can be made to ``useful'' values. For example, the age of a person cannot be a negative value in the real world, and it is unrealistic that a person would be more than 150 years old. Therefore, we can impose a range of age values that we might try. To test error handling for negative ranges, we might set the minimum value to $-1$. This allows the metaheuristic to try a negative value, while preventing it from wasting time trying \textit{many} negative values.

In this example, we assume that a tester would create this metadata file---a task that would take only a few minutes for a UUT. However, it would be possible to write code to extract this information as well.

\smallskip\noindent\textbf{Random Test Generation:} Both of the presented metaheuristic algorithms start by making random ``guesses''---either generating random test cases or generating entire test suites at random---and will occasionally modify solutions through random generation of additional elements. To control the size of the generated test suites or test cases, there are two user-controllable parameters:
\begin{itemize}
    \item \textbf{Maximum number of test cases:} The largest test suite that can be randomly generated. When a suite is generated, a size is chosen between 1 - \pythoninline{max_test_cases}, and that number of test cases are generated and added to the suite.
    \item \textbf{Maximum number of actions:} The largest individual test case that can be randomly generated. When a test case is generated, a number of actions between 1 - \pythoninline{max_actions} is chosen and that many actions are added to the test case (following a constructor call). 
\end{itemize}
\noindent By default, we use 20 as the value for both parameters. This provides a reasonable starting point for covering a range of interesting behaviors, while preventing test suites from growing large enough to hinder debugging. Test suites can then grow or shrink over time through manipulation by the metaheuristic.

\smallskip\noindent\textbf{Search Budget:} This search budget is the time allocated to the metaheuristic. The goal of the metaheuristic is to find the best solution possible within this limitation. This parameter is also user-controlled:
\begin{itemize}
    \item \textbf{Search Budget:} The maximum number of generations of work that can be completed before returning the best solution found.
\end{itemize}
\noindent The search budget is expressed as a number of \textit{generations}---cycles of exploration of the search space of test inputs---that are allocated to the algorithm. This can be set according to the schedule of the tester. By default, we allow 200 generations in this example. However, fewer may still produce acceptable results, while more can be allocated if the tester is not happy with what is returned in that time frame.

\subsubsection{Hill Climber}

\begin{figure}
\centering
\small
\begin{minted}{python}
# Generate an initial random solution, and calculate its fitness
solution_current = Solution()
solution_current.test_suite = generate_test_suite(metadata, max_test_cases, max_actions)
calculate_fitness(metadata, fitness_function, num_tests_penalty,
                  length_test_penalty, solution_current)

# The initial solution is the best we have seen to date
solution_best = copy.deepcopy(solution_current)

# Continue to evolve until the generation budget is exhausted 
# or the number of restarts is exhausted.
gen = 1
restarts = 0

while gen <= max_gen and restarts <= max_restarts:
    tries = 1
    changed = False

    # Try random mutations until we see a better solutions, 
    # or until we exhaust the number of tries.
    while tries < max_tries and not changed:
        solution_new = mutate(solution_current)
        calculate_fitness(metadata, fitness_function, num_tests_penalty,
                          length_test_penalty, solution_new)

        # If the solution is an improvement, make it the new solution.
        if solution_new.fitness > solution_current.fitness:
            solution_current = copy.deepcopy(solution_new)
            changed = True

            # If it is the best solution seen so far, then store it.
            if solution_new.fitness > solution_best.fitness:
                solution_best = copy.deepcopy(solution_current)

        tries += 1
        
    # Reset the search if no better mutant is found within a set number 
    # of attempts by generating a new solution at random.
    if not changed:
        restarts += 1
        solution_current = Solution()
        solution_current.test_suite = generate_test_suite(metadata, max_test_cases, 
                                                          max_actions)
        calculate_fitness(metadata, fitness_function, num_tests_penalty,
                          length_test_penalty, solution_current)

    # Increment generation
    gen += 1

# Return the best suite seen
\end{minted}
\caption{The core body of the Hill Climber algorithm.}
\label{fig:hillclimber}
\end{figure}

A Hill Climber is a classic metaheuristic that embodies the ``guess-and-check'' process we discussed earlier. The algorithm makes an initial guess purely at random, then attempts to improve that guess by making small, iterative changes to it. When it lands on a guess that is better than the last one, it adopts it as the current solution and proceeds to make small changes to that solution. The core body of this algorithm is shown in Figure~\ref{fig:hillclimber}. The full code can be found at \url{https://github.com/Greg4cr/PythonUnitTestGeneration/blob/main/src/hill_climber.py}.

The variable \pythoninline{solution_current} stores the current solution. At first, it is initialised to a random test suite, and we measure the fitness of the solution (lines 2-5). Following this, we start our first generation of evolution. While we have remaining search budget, we then attempt to improve the current solution.

Each generation, we attempt to improve the current solution through the process of \textit{mutation}. During mutation, we introduce a small change to the current solution. Below, we outline the types of change possible during mutation: 

\begin{center}
\includegraphics[width=0.8\textwidth]{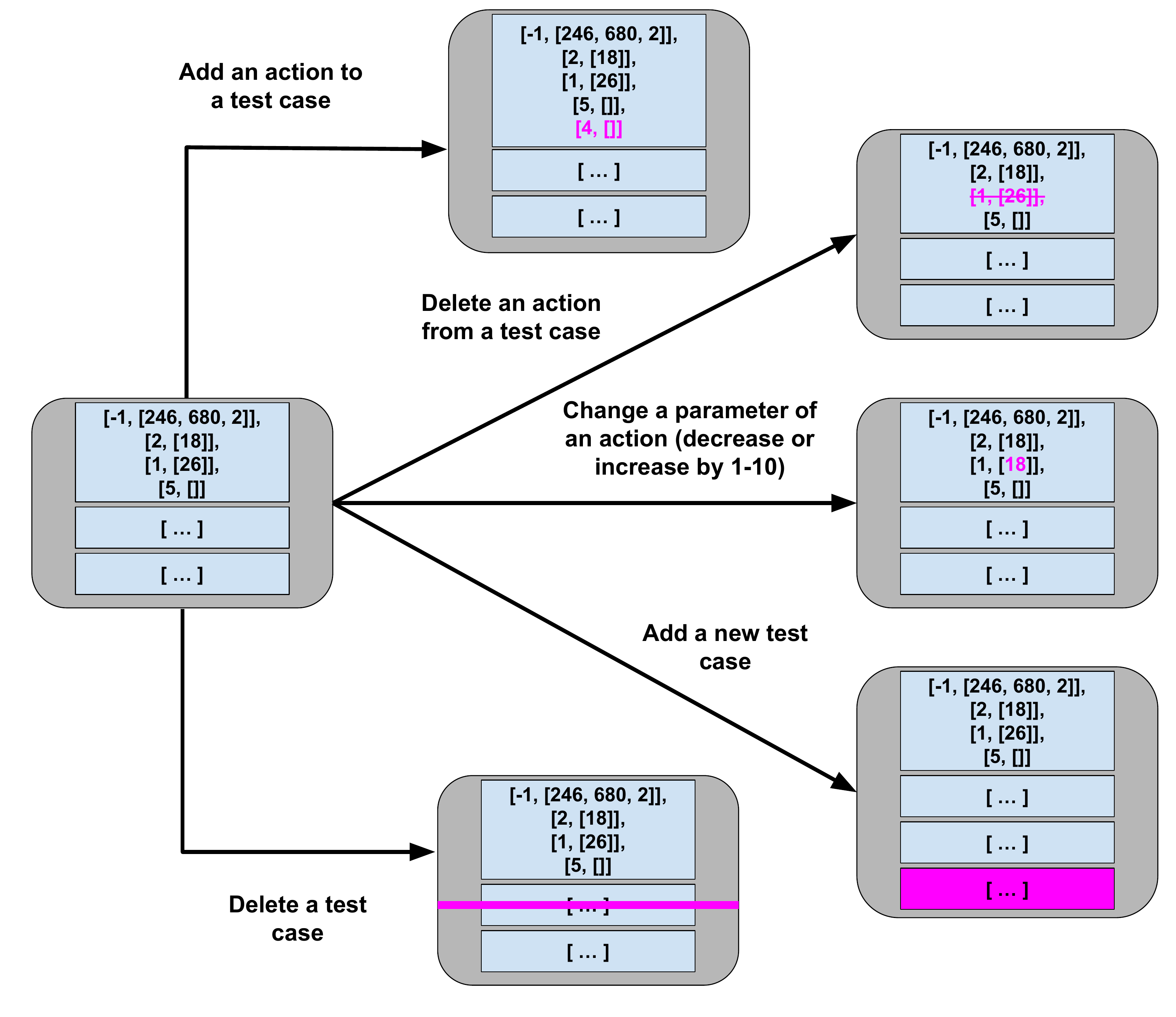}
\end{center}

After selecting and applying one of these transformations (line 22), we measure the fitness of the mutated solution (line 23). It if it better than the current solution, we make the mutation into the current solution (lines 26-28). If it is better than the best solution seen to date, we also save it as the new best solution (lines 30-31). We then proceed to the next generation. 

If the mutation is not better than the current solution, we try a different mutation to see if it is better. The range of transformations results in a very large number of possible transformations. However, even with such a range, we may end up in situations where no improvement is possible, or where it would be prohibitively slow to locate an improved solution. We refer to these situations at \textit{local optima}---solutions that, while they may not be the best possible, are the best that can be located through incremental changes. 

We can think of the landscape of possible solutions as a topographical map, where better fitness scores represent higher levels of elevation in the landscape. This algorithm is called a ``Hill Climber'' because it attempts to scale that landscape, finding the tallest peak that it can in its local neighborhood. 

\begin{figure}
\centering
\small
\begin{minted}{python}
#Create initial population.
population = create_population(population_size)

#Initialise best solution as the first member of that population.
solution_best = copy.deepcopy(population[0])

# Continue to evolve until the generation budget is exhausted.
# Stop if no improvement has been seen in some time (stagnation).
gen = 1
stagnation = -1

while gen <= max_gen and stagnation <= exhaustion:
    # Form a new population.
    new_population = []

    while len(new_population) < len(population):
        # Choose a subset of the population and identify 
        # the best solution in that subset (selection).
        offspring1 = selection(population, tournament_size)
        offspring2 = selection(population, tournament_size)

        # Create new children by breeding elements of the best solutions 
        # (crossover).
        if random.random() < crossover_probability:
            (offspring1, offspring2) = uniform_crossover(offspring1, offspring2)

        # Introduce a small, random change to the population (mutation).
        if random.random() < mutation_probability:
            offspring1 = mutate(offspring1)
        if random.random() < mutation_probability:
            offspring2 = mutate(offspring2)

        # Add the new members to the population.
        new_population.append(offspring1)
        new_population.append(offspring2)

        # If either offspring is better than the best-seen solution, 
        # make it the new best.
        if offspring1.fitness > solution_best.fitness:
            solution_best = copy.deepcopy(offspring1)
            stagnation = -1
        if offspring2.fitness > solution_best.fitness:
            solution_best = copy.deepcopy(offspring2)
            stagnation = -1

    # Set the new population as the current population.
    population = new_population

    # Increment the generation.
    gen += 1
    stagnation += 1

# Return the best suite seen
\end{minted}
\caption{The core body of the Genetic Algorithm.}
\label{fig:genetic_algorithm} \vspace{-5pt}
\end{figure}

If we reach a local optima, we need to move to a new ``neighborhood'' in order to find taller peaks to ascend. In other words, when we become stuck, we restart by replacing the current solution with a new random solution (lines 37-42). Throughout this process, we track the best solution seen to date to return at the end. To control this process, we use two user-controllable parameters.
\begin{itemize}
    \item \textbf{Maximum Number of Tries:} A limit on the number of mutations we are willing to try before restarting (\pythoninline{max_tries}, line 21). By default, this is set to 200.
    \item \textbf{Maximum Number of Restarts:} A limit of restarts we are willing to try before giving up on the search (\pythoninline{max_restarts}, line 15). Be default, this is set to 5.
\end{itemize}

The core process employed by the Hill Climber is simple, but effective. Hill Climbers also tend to be faster than many other metaheuristics. This makes them a popular starting point for search-based automation. Their primary weakness is their reliance on making a good initial guess. A bad initial guess could result in time wasted exploring a relatively ``flat'' neighborhood in that search landscape. Restarts are essential to overcoming that limitation. 

\subsubsection{Genetic Algorithm}

Genetic Algorithms model the evolution of a population over time. In a population, certain individuals may be ``fitter'' than others, possessing traits that lead them to thrive---traits that we would like to see passed forward to the next generation through reproduction with other fit individuals. Over time, random mutations introduced into the population may also introduce advantages that are also passed forward to the next generation. Over time, through mutation and reproduction, the overall population will grow stronger and stronger. 

As a metaheuristic, a Genetic Algorithm is build on a core generation-based loop like the Hill Climber. However, there are two primary differences:
\begin{itemize}
    \item Rather than evolving a single solution, we simultaneously manage a \textit{population} of different solutions. 
    \item In addition to using mutation to improve solutions, a Genetic Algorithm also makes use of a \textit{selection} process to identify the best individuals in a population, and a \textit{crossover} process that produces new solutions merging the test cases (``genes'') of parent solutions (``chromsomes'').
\end{itemize}

The core body of the Genetic Algorithm is listed in Figure~\ref{fig:genetic_algorithm}. The full code can be found at \url{https://github.com/Greg4cr/PythonUnitTestGeneration/blob/main/src/genetic_algorithm.py}.

We start by creating an initial population, where each member of the population is a randomly-generated test suite (line 1). We initialise the best solution to the first member of that population (line 5). We then begin the first generation of evolution (line 12). 

Each generation, we form a new population by applying a series of actions intended to promote the best ``genes'' forward. We form the new population by creating two new solutions at a time (line 16). First, we attempt to identify two of the best solutions in a population. If the population is large, this can be an expensive process. To reduce this cost, we perform a \textit{selection} procedure on a randomly-chosen subset of the population (lines 19-20), explained below:

\begin{center}
\includegraphics[width=0.75\textwidth]{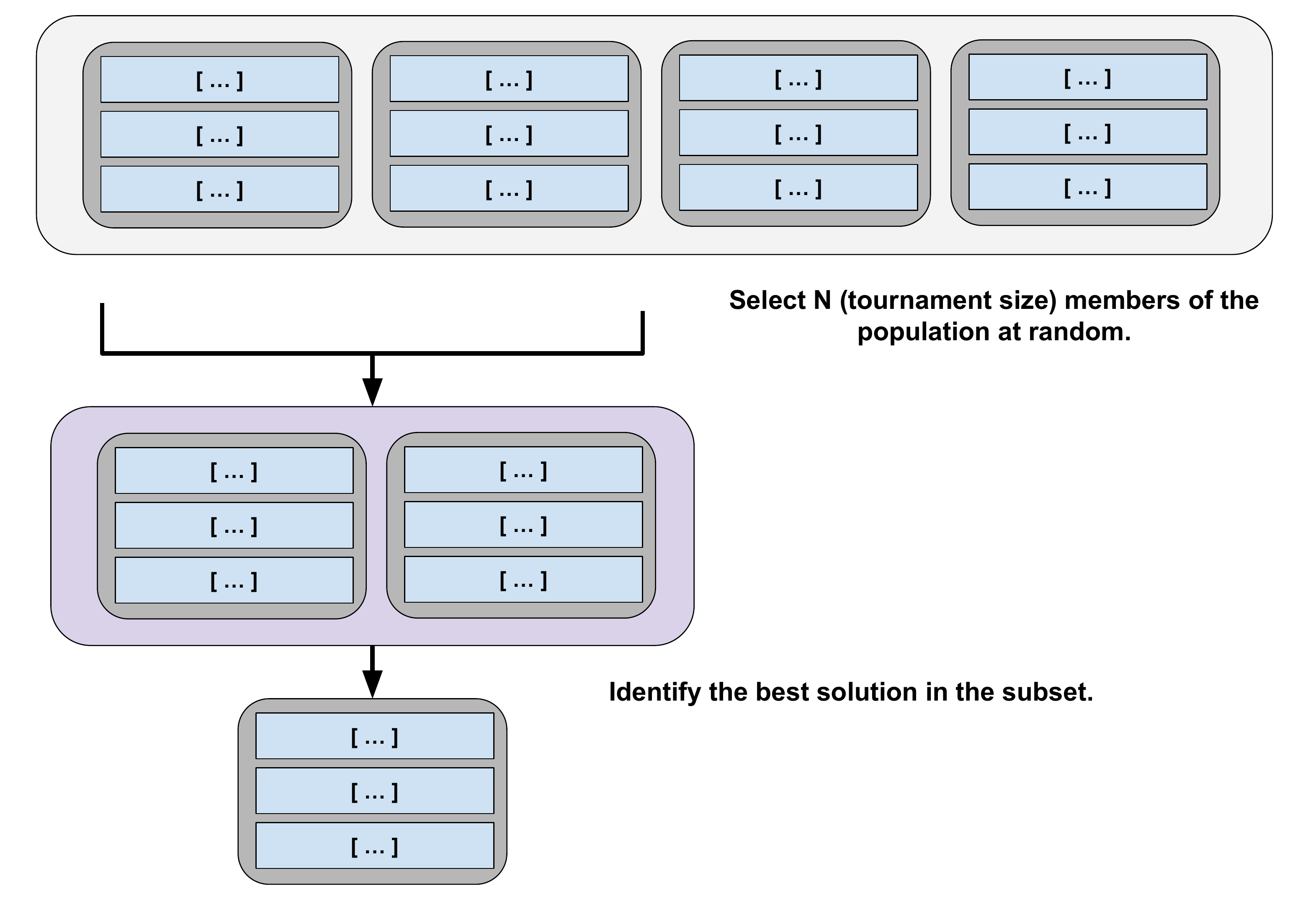} 
\end{center}

The fitness of the members of the chosen subset is compared in a process called ``tournament'', and a winner is selected. The winner may not be the best member of the full population, but will be at least somewhat effective, and will be identified at a lower cost than comparing all population members. These two solutions may be carried forward as-is. However, at certain probabilities, we may make further modifications to the chosen solutions.

The first of these is crossover---a transformation that models reproduction. We generate a random number and check whether it is less than a user-set \pythoninline{crossover_probability} (line 23). If so, we combine individual genes (test cases) of the two solutions using the following process: 

\begin{center}
\includegraphics[width=0.8\textwidth]{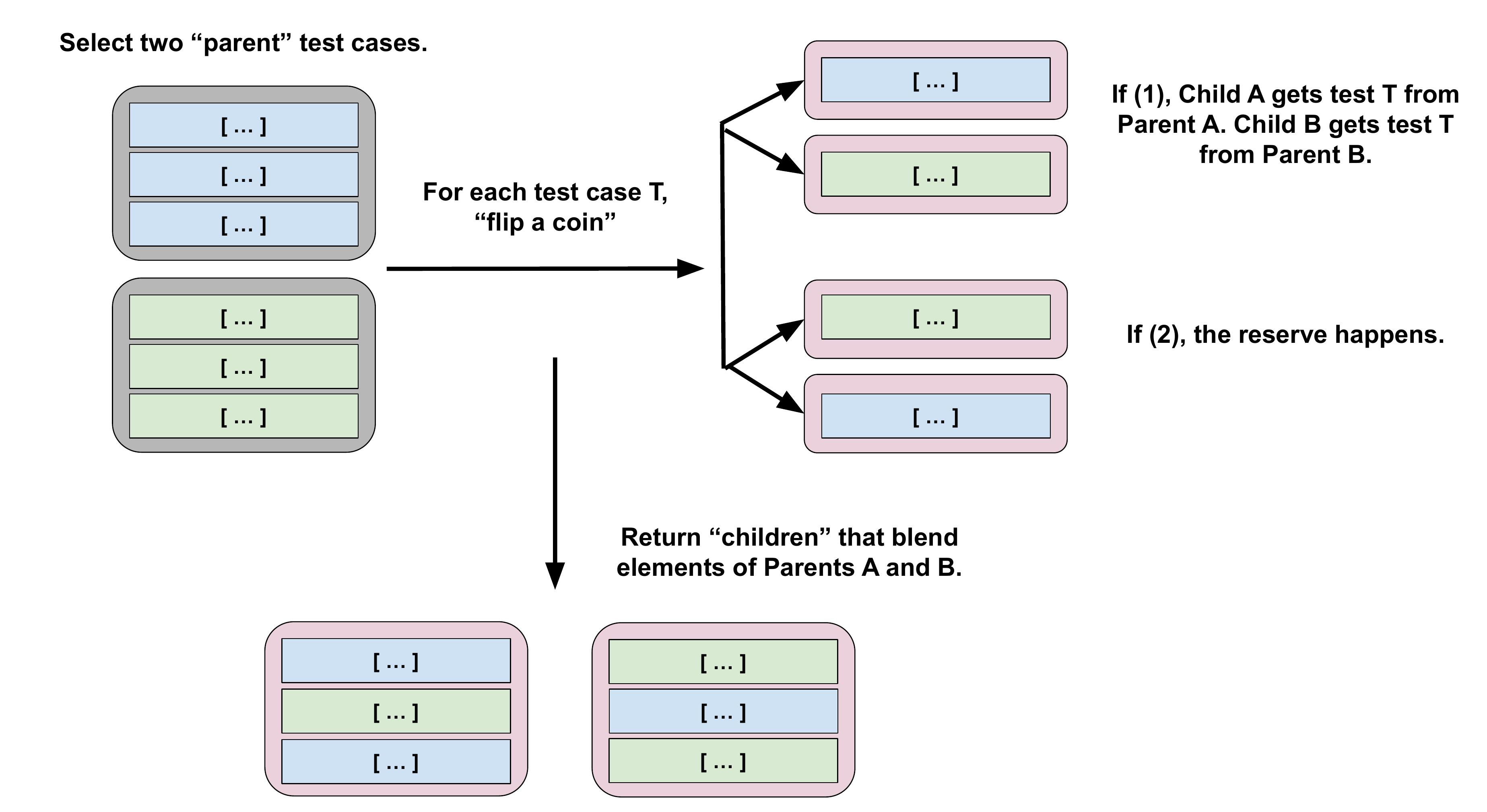} 
\end{center}

If the parents do not contain the same number of test cases, then the remaining cases can be randomly distributed between the children. This form of crossover is known as ``uniform crossover''. There are other means of performing crossover. For example, in ``single-point'' crossover, a single index is chosen, and one child gets all elements from Parent A before that index, and all elements from Parent B from after that index (with the other child getting the reverse). Another form, ``discrete recombination'', is similar to uniform crossover, except that we make the coin flip for each child instead of once for both children at each index.

We may introduce further mutations to zero, one, or both of the solutions. If a random number is less than a user-set \pythoninline{mutation probability} (lines 27, 29), we will introduce a single mutation to that solution. We do this independently for both solutions. The mutation process is the same as in the Hill Climber, where we can add, delete, or modify an individual action or add or delete a full test case. 

Finally, we add both of the solutions to the new population (line 46). If either solution is better than the best seen to date, we save it to be returned at the end of the process (lines 38-43). Once the new population is complete, we continue to the next generation.

There may be a finite amount of improvement that we can see in a population before it becomes \textit{stagnant}. If the population cannot be improved further, we may wish to terminate early and not waste computational effort. To enable this, we count the number of generations where no improvement has been seen (line 50), and terminate if it passes a user-set \textit{exhaustion} threshold (line 12). If we identify a new ``best'' solution, we reset this counter (lines 40, 43). 

The following parameters of the genetic algorithm can be adjusted:
\begin{itemize}
    \item \textbf{Population Size:} The size of the population of solutions. By default, we set this to 20. This size must be a even number in the example implementation.
    \item \textbf{Tournament Size:} The size of the random population subset compared to identify the fittest population members. By default, this is set to 6.
    \item \textbf{Crossover Probability:} The probability that we apply crossover to generate child solutions. By default, $0.7$. 
    \item \textbf{Mutation Probability:} The probability that we apply mutation to manipulate a solution. By default, $0.7$.
    \item \textbf{Exhaustion Threshold:} The number of generations of stagnation allowed before the search is terminated early. By default, we have set this to 30 generations.
\end{itemize}

These parameters can have a noticeable effect on the quality of the solutions located. Getting the best solutions quickly may require some experimentation. However, even at default values, this can be a highly effective method of generating test suites.

\subsection{Examining the Resulting Test Suites}\label{sec:results} 

\begin{figure}[!t]
\centering
   \begin{subfigure}[t]{0.5\textwidth}
        \centering
        \includegraphics[width=\textwidth]{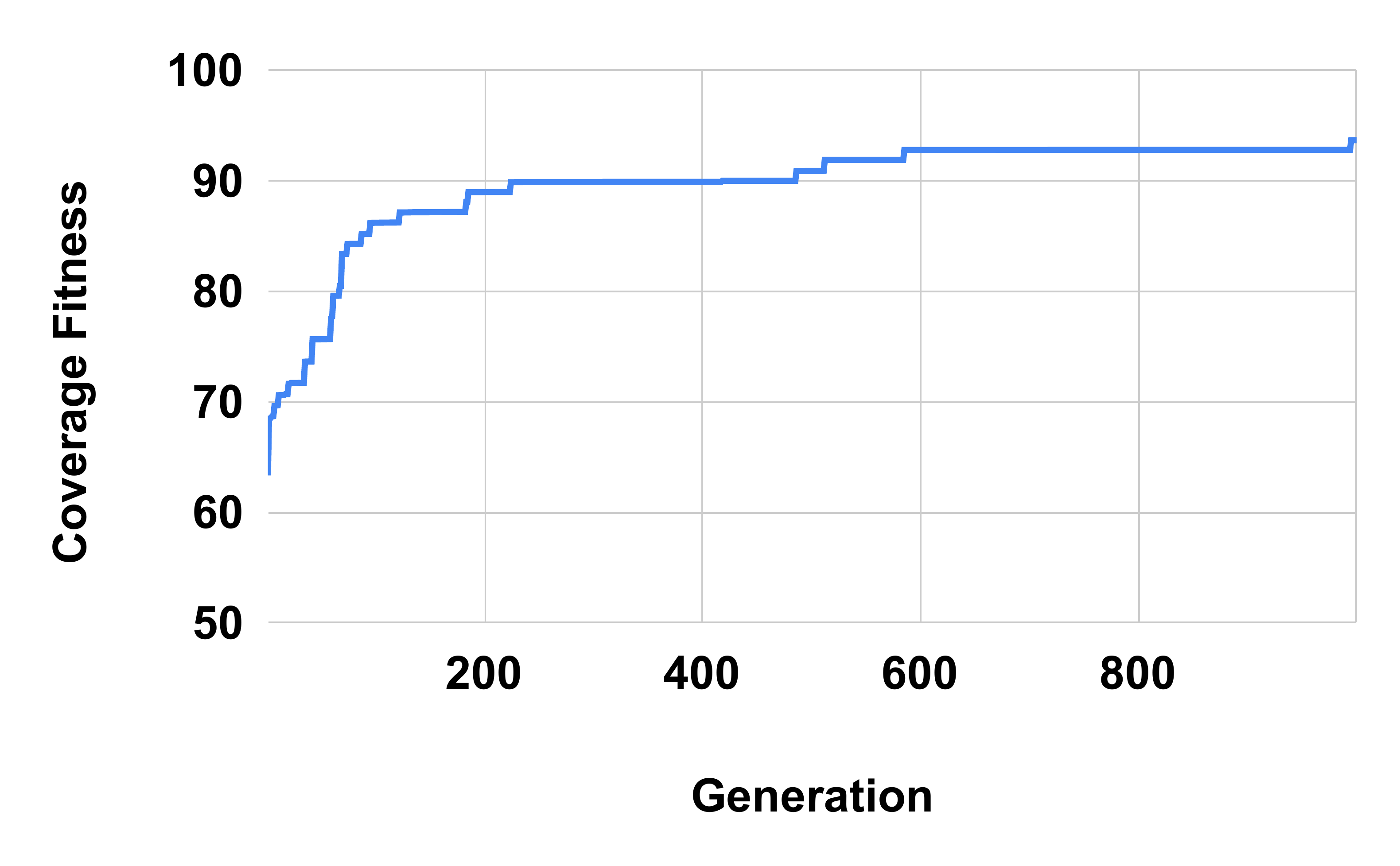} \vspace{-10pt}
    \end{subfigure}%
       \begin{subfigure}[t]{0.5\textwidth}
        \centering
        \includegraphics[width=\textwidth]{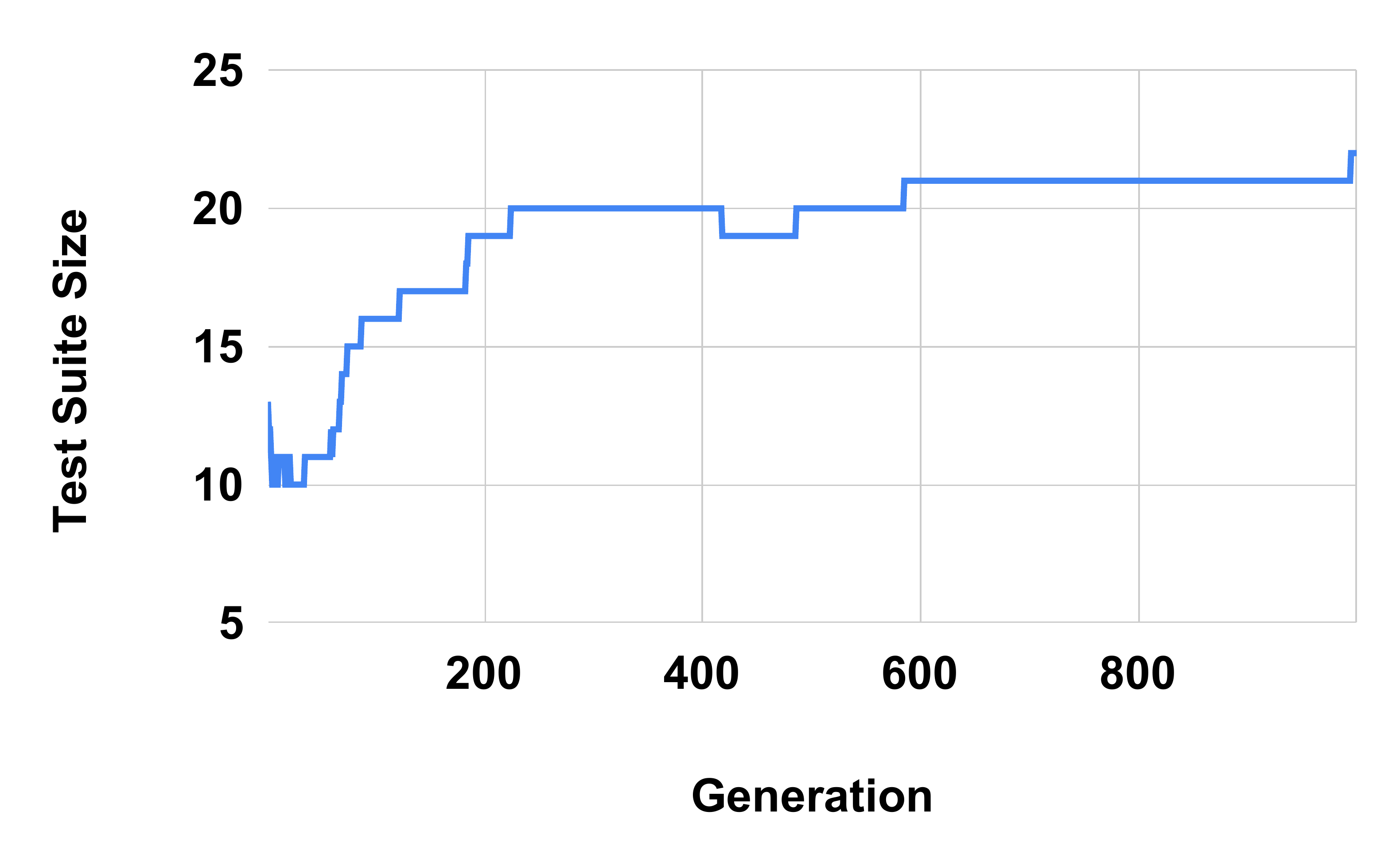} \vspace{-10pt}
    \end{subfigure}%
    
    \begin{subfigure}[t]{0.5\textwidth}
        \centering
        \includegraphics[width=\textwidth]{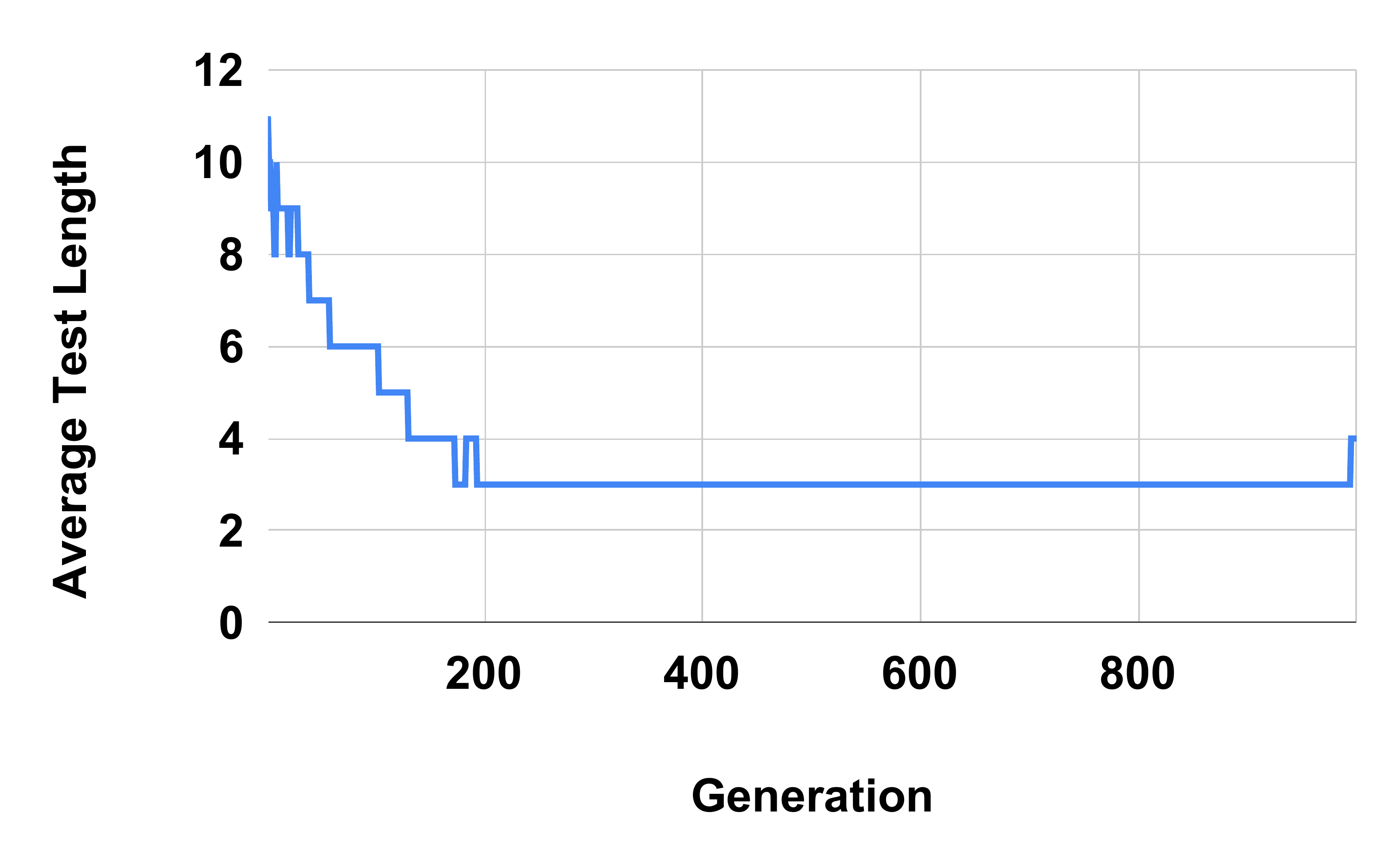} \vspace{-10pt}
    \end{subfigure}%
\caption{Change in fitness, test suite size, and average number of actions in a test case over 1000 generations. Note that fitness includes both coverage and bloat penalty, and can never reach 100.}
\label{fig:results} 
\end{figure}

Now that we have all of the required components in place, we can generate test suites and examine the results. To illustrate what these results look like, we will examine test suites generated after executing the Genetic Algorithm for 1000 generations. During these executions, we disabled the exhaustion threshold to see what would happen if the algorithm was given the full search budget to work. 

Figure~\ref{fig:results} illustrates the results of executing the Genetic Algorithm. We can see the change in fitness over time, as well as the change in the number of test cases in the suite and the average number of actions in test cases. Note that fitness is penalised by the bloat penalty, so the actual statement coverage is higher than the final fitness value. Also note that metaheuristic search algorithms are random. Therefore, each execution of the Hill Climber or Genetic Algorithm will yield different test suites in the end. Multiple executions may be desired in order to detect additional crashes or other issues.

The fitness starts around 63, but quickly climbs until around generation 100, when it hits approximately 86. There are further gains after that point, but progress is slow. At generation 717, it hits a fitness value of 92.79, where it remains until near the very end of the execution. At generation 995, a small improvement is found that leads to the coverage of additional code and a fitness increase to 93.67. Keep in mind, again, that a fitness of ``100'' is not possible due to the bloat penalty. It is possible that further gains in fitness could be attained with an even higher search budget, but covering the final statements in the code and further trimming the number or length of test cases both become quite difficult at this stage. 

The test suite size starts at 13 tests, then sheds excess tests for a quick gain in fitness. However, after that, the number of tests rises slowly as coverage increases. For much of the search, the test suite remains around 20 test cases, then 21. At the end, the final suite has 22 test cases. In general, it seems that additional code coverage is attained by generating new tests and adding them to the suite. 

At times, redundant test cases are removed, but instead, we often see redundancy removed through the deletion of actions within individual test cases. The initial test cases are often quite long, with many redundant function calls. Initially, tests have an average of 11 actions. Initially, the number of actions oscillates quite a bit between an average of 8-10 actions. However, over time, the redundant actions are trimmed from test cases. After generation 200, test cases have an average of only three actions until generation 995, when the new test case increases the average length to four actions. With additional time, it is likely that this would shrink back to three. We see that the tendency is to produce a large number of very small test cases. This is good, as short test cases are often easier to understand and make it easier to debug the code to find faults. 

\begin{figure}
\centering
\small
\begin{minted}{python}
def test_0():
    cut = bmi_calculator.BMICalc(120,860,13)
	cut.classify_bmi_teens_and_children()

def test_2():
	cut = bmi_calculator.BMICalc(43,243,59)
	cut.classify_bmi_adults()
	cut.height = 526
	cut.classify_bmi_adults()
	cut.classify_bmi_adults()

def test_5():
	cut = bmi_calculator.BMICalc(374,343,17)
	cut.age = 123
	cut.classify_bmi_adults()
	cut.age = 18
	cut.classify_bmi_teens_and_children()
	cut.weight = 396
	cut.classify_bmi_teens_and_children()

def test_7():
	cut = bmi_calculator.BMICalc(609,-1,94)

def test_11():
	cut = bmi_calculator.BMICalc(491,712,20)
	cut.classify_bmi_adults()

def test_17():
	cut = bmi_calculator.BMICalc(608,717,6)
	cut.classify_bmi_teens_and_children()
	cut.age = 91
	cut.classify_bmi_teens_and_children()
	cut.classify_bmi_teens_and_children()

	\end{minted}
\caption{A subset of the test suite produced by the Genetic Algorithm, targeting statement coverage.}
\label{fig:stmt_suite} 
\end{figure}

More complex fitness functions or algorithms may be able to cover more code, or cover the same code more quickly, but these results show the power of even simple algorithms to generate small, effective test cases.
A subset of a final test suite is shown in Figure~\ref{fig:stmt_suite}.\footnote{The full suite can be found at \url{https://github.com/Greg4cr/PythonUnitTestGeneration/blob/main/src/example/test_bmi_calculator_automated_statement.py}.}

Some test cases look like \pythoninline{test_0()} and \pythoninline{test_11()} in the example---a constructor call, followed by a BMI classification. Others will adjust the variable assignments, then make calls. For example, \pythoninline{test_5()} covers several paths in the code by making assignments, then getting classifications, multiple times. \pythoninline{test_7()} is an example of one where only a constructor call was needed, as the value supplied---a negative weight, in this case---was sufficient to trigger an exception. 

There is still some room for improvement in these test cases. For example, \pythoninline{test_2()} and \pythoninline{test_17()} both contain redundant calls to a classification method. It is likely that a longer search budget would remove these calls. It would be simple to simply remove all cases where a method is called twice in a row with the same arguments from the suite. However, in other cases, those calls may have different results (e.g., if class state was modified by the calls), and you would want to leave them in place.

Search-based test generation requires a bit of experimentation. Even the Hill Climber has multiple user-selectable parameters. Finding the right search budget, for example, can require some experimentation. It may be worth executing the algorithm once with a very high search budget in order to get an initial idea of the growth in fitness. In this case, a tester could choose to stop much earlier than 1000 generations with little loss in effectiveness. For example, only limited gains are seen after 200 generations, and almost no gain in fitness is seen after 600 generations.

\subsection{Assertions}
 
It is important to note that this chapter is focused on \textit{test input} generation. These test cases lack assertion statements, which are needed to check the correctness of the program behavior. 

These test cases \textit{can} be used as-is. Any exceptions thrown by the UUT, or other crashes detected, when the tests execute will be reported as failures. In some cases, exceptions \textit{should} be thrown. In Figure~\ref{fig:stmt_suite}, \pythoninline{test_17} will trigger an exception when \pythoninline{classify_bmi_teens_and_children()} is called for a 91-year-old. This exception is the desired behavior. However, in many cases, exceptions are not desired, and these test cases can be used to alert the developer about crash-causing faults. 

Otherwise, the generated tests will need assertions to be added. A tester can add assertions manually to these test cases, or a subset of them, to detect incorrect output. Otherwise, researchers and developers have begun to explore the use of AI techniques to generate assertions as well. We will offer pointers to some of this work in Section~\ref{sec:advanced}. 

%% file: advanced.tex
The input generation technique introduced in the previous section can be used to generate small, effective test cases for Python classes. This section briefly introduces concepts that build on this foundation, and offers pointers for readers interested in developing more complex AI-enhanced unit test automation tools.

\subsection{Distance-Based Coverage Fitness Function}

This chapter introduced the idea that we can target the maximization of code coverage as a fitness function. We focused on  statement coverage---a measurement of the number of lines of code executed. A similar measurement is the \textit{branch coverage}---a measurement of the number of outcomes of control-altering expressions covered by test cases. This criterion is expressed over statements that determine which code will be executed next in the sequence. For example, in Python, this includes \pythoninline{if}, \pythoninline{for}, and \pythoninline{while} statements. Full branch coverage requires \pythoninline{True} and \pythoninline{False} outcomes for the Boolean predicates expressed in each statement. 

Branch coverage can be maximised in the same manner that we maximised statement coverage---by simply measuring the attained coverage and favoring higher totals. However, advanced search-based input generation techniques typically use a slightly more complex fitness function based on \textit{how close} a test suite came to covering each of these desired outcomes. 

Let's say that we had two test suites---one that attains 50\% branch coverage and one that attains 75\% coverage. We would favor the one with 75\% coverage, of course. However, what if both had 75\% coverage? Which is better? The answer is that we want the one that is \textit{closer} to covering the remaining 25\%. Perhaps, with only small changes, that one could attain 100\% coverage. We cannot know which of those two is better with our simple measurement of coverage. Rather, to make that determination, we divide branch coverage into a set of \textit{goals}, or combinations of an expression we want to reach and an outcome we desire for that expression. Then, for each goal, we measure the \textit{branch distance} as a score ranging from $0-1$. The branch distance is defined as follows:
\begin{equation}
\scriptsize
    {distance}(goal, suite) = 
    \begin{cases}
    0 & \text{If the branch is reached and the desired outcome is attained.} \\
    {distance}_{min}(goal, suite) & \text{If the branch is reached, but the desired outcome is not attained.} \\
    1 & \text{If the branch has not been reached.}
    \end{cases}
\end{equation}

Our goal is to \textit{minimise} the branch distance. If we have \textit{reached} the branch of interest and \textit{attained} the desired outcome, then the score is 0. If we have not reached the branch, then the value is 1. If we have reached the branch, but not covered it, then we measure \textit{how close} we came by transforming the Boolean predicate into a numeric function. For example, if we had the expression \pythoninline{if x == 5:} and desired a \pythoninline{True} outcome, but \pythoninline{x} was assigned a value of 3 when we executed the expression, we would calculate the branch distance as $abs(x - 5) = abs(3 - 5) = 2$.\footnote{For more information on this calculation, and normalization, see the explanations from McMinn, Lukasczyk, and Arcuri:~\cite{McMinn04:SBTesting,Lukascyzk20:Python,Arcuri13:Normalize}.} 

We then normalise this value to be between 0 and 1. As this expression may be executed multiple times by the test suite, we take the minimum branch distance as the score. We can then attain a fitness score for the test suite by taking the sum of the branch distances for all goals: $fitness = \sum_{{goal} \in {Goals}} distance(goal, suite)$. 

The branch distance offers a fine-grained score that is more informative than simply measuring the coverage. Using this measurement allows faster attainment of coverage, and may enable the generation tool to attain more coverage than would otherwise be possible. The trade-off is the increased complexity of the implementation. At minimum, the tool would have to insert logging statements into the program. To avoid introducing side-effects into the behavior of the class-under-test, measuring the branch distance may require complex instrumentation and execution monitoring.

\subsection{Multiple and Many Objectives}



When creating test cases, we typically have many different goals. A critical goal is to cover all important functionality, but we also want few and short test cases, we want tests to be understandable by humans, we want tests to have covered all parts of the system code, and so on. When you think about it, it is not uncommon to come up with five or more goals you have during test creation. If we plan to apply AI and optimisation to help us to create these test cases, we must encode these goals so that they are quantitative and can be automatically and quickly checked. We have the ability to do this through fitness functions. However, if we have multiple goals, we cannot settle for single-objective optimisation and instead have to consider the means to optimise all of these objectives at the same time.

A simple solution to the dilemma is to try to merge all goals together into a single fitness function which can then be optimised, often by adding all functions into a single score---potentially weighting each. For example, if our goals are high code coverage and few test cases, we could normalise the number of uncovered statements and the number of test cases to the same scale, sum them, and attempt to minimise this sum. 

However, it is almost inevitable that many of your goals will compete with each other. In this two-objective example, we are punished for adding more test cases, but we are also punished if we do not cover all code. If these two goals were considered equally important, it seems possible that an outcome could be a single, large test case that tries to cover as much of the code as possible. While this might be optimal given the fitness function we formulated, it might not reflect what you really hope to receive from the generation tool. In general, it will be very hard to decide up-front how you want to trade off one objective versus the others. Even if you can in principle set weights for the different elements of the fitness function, when the objectives are fundamentally at odds with each other, there is no single weight assignment that can address all conflicts.

An alternative, and often better, solution is to keep each fitness function separate and attempt to optimise all of them at the same time, balancing optimisation of one with optimisation of each of the others. The outcome of such a multi-objective optimisation is not a single best solution, but a set of solutions that represent good trade-offs between the competing objectives. The set approximates what is known as the Pareto frontier, which is the set of all solutions that are not dominated by any other solution. A solution dominates another one if it is at least as good in all the objectives and better in at least one. This set of solutions represents balancing points, where the solution is the best it can be at some number of goals without losing attainment of the other goals. In our two-objective example of code coverage and test suite size, we might see a number of solutions with high coverage and a low number of test cases along this frontier, with some variation representing different trade-offs between these goals. We could choose the solution that best fits our priorities---perhaps taking a suite with 10 tests and 93\% coverage over one with 12 tests and 94\% coverage.


One well-known example of using multi-objective optimisation in software testing is the Sapienz test generation developed by Facebook to test Android applications through their graphical user interface~\cite{mao2016sapienz}. It can generate test sequences of actions that maximise code coverage and the number of crashes, while minimizing the number of actions in the test cases. The system, thus, simultaneously optimises three different objectives. It uses a popular genetic algorithm known as NSGA-II for multi-objective optimisation and returns a set of non-dominated test cases.

When the number of objectives grows larger, some of the more commonly used optimisation algorithms---like NSGA-II---become less effective. Recently, ``many-objective'' optimisation algorithms that are more suited to such situations have been proposed. One such algorithm was recently used to select and prioritise test cases for testing software product lines~\cite{hierons2002comparing}. A total of nine different fitness functions are optimised by the system. In addition to the commonly used test case and test suite sizes, other objectives included are the pairwise coverage of features, dissimilarity of test cases, as well as the number of code changes the test cases cover.

\subsection{Human-readable Tests}


A challenge with automated test generation is that the generated test cases typically do not look similar to test cases that human developers and testers would write. Variable names are typically not informative and the ordering of test case steps might not be natural or logical for a human reading and interpreting them. This can create challenges for using the generated test cases. Much of existing research on test generation has not considered this a problem. A common argument has been that since we can generate so many test cases and then automatically run them there is little need for them to be readable; the humans will not have the time or interest to analyse the many generated test cases anyway.
However, in some scenarios we really want to generate and then keep test cases around, for example when generating test cases to reach a higher level of code coverage. Also, when an automatically generated test case fails it is likely that a developer will want to investigate its steps to help identify what leads the system to fail. Automated generation of readable test cases would thus be helpful.

One early result focused on generating XML test inputs that were more comprehensible to human testers~\cite{poulding2015automated}. The developed system could take any XSD (XML Schema Definition) file as input and then create a model from which valid XML could then be generated. A combination of several AI techniques were then used to find XML inputs that were complex enough to exercise the system under test enough but not too complex since that would make the generated inputs hard for humans to understand. Three different metrics of complexity was used for each XML inputs (its number of elements, attributes, and text nodes) and the AI technique of Nested Monte-Carlo Search, an algorithm very similar to what was used in the AlphaGO Go playing AI~\cite{silver2016mastering},
were then used to find good inputs for them. Results were encouraging but it was found that not all metrics were as easily optimised by the chosen technique. Also, for real comprehensibility it will not be enough to only find the right size of test inputs; the specific content and values in them will also be critical. 

Other studies have found that readability can be increased by---for example---using real strings instead of random ones (e.g., by pulling string values from documentation), inserting default values for ``unimportant'' elements (rather than omitting them), and limiting the use and mixture of \pythoninline{null} values with normal values\footnote{Although, of course, some \pythoninline{null} values should be applied to catch common ``null pointer'' faults.}~\cite{McMinn10:humanoracle,Alsharif2019:HOC-SQL}.

A more recent trend in automated software engineering is to use techniques from the AI area of natural language processing on source code. For example, GitHub in 2021 released its Co-Pilot system, which can auto-complete source code while a developer is writing it~\cite{chen2021evaluating}. They used a neural network model previously used for automatically generating text that look like it could have been written by humans. Instead of training it on lots of human-written texts they instead trained it on human-written source code. The model can then be used to propose plausible completions of the source code currently being written by a developer in a code editor. In the future, it is likely that these ideas can and will also be used to generate test code. However, there are many risks with such approaches, and it is not a given that the generated test code will be meaningful or useful in actual testing. For example, it has been shown that Co-Pilot can introduce security risks~\cite{pearce2021empirical}. Still, by combining these AI techniques with multi-objective optimisation it seems likely that we can automatically generate test cases that are both useful and understandable by humans.

\subsection{Finding Input Boundaries}


One fundamental technique to choose test input is known as boundary value testing\slash analysis. This technique aims to identify input values at the \textit{boundary} between different visible program behaviours, as those boundaries often exhibit faults due to---for example---``off-by-one'' errors or other minor mistakes. Typically, testers manually identify boundaries by using the software specification to define different \textit{partitions}, i.e., sets of input that exhibit similar behaviours. Consider, for example, the creation of date objects. Testers can expect that valid days mainly lie within the range of 1--27. However, days greater or equal than 28 might reveal different outputs depending on the value chosen for month or year (e.g., February 29th). Therefore, most testers would choose input values between 28--32 as \textit{one of} the boundaries for testing both valid and invalid dates (similarly, boundary values for day between 0--1).

The \textit{program derivative} measures the program's sensitivity to behavioural changes for different sets of input values~\cite{felt2020derivative}. Analogous to the mathematical concept of a derivative, the program derivative conveys how function values (output) change when varying an independent variables (input). In other words, we can detect boundary values by detecting notable output differences when using a similar sets of inputs~\cite{dobslaw2020bve}. We quantify the similarities between input and output by applying various distance functions that quantify the similarity between a pair of values. Low distance values indicate that the pair of values are similar to each other. Some of the widely used distance functions are the Jaccard index (strings), Euclidean distance (numerical input) or even the more generic Normalised Compressed Distance (NCD). 

The program derivative analyses the ratio between the distances of input and output of a program under test (Equation X). Let $a$ and $b$ be two different input values for program $P$ with corresponding output values $P(a)$ and $P(b)$. We use the distance functions $d_i(a,b)$ and $d_o(P(a), P(b))$ to measure the distance between, respectively, the pair of input and their corresponding output values. The program derivative (PD) is defined as~\cite{dobslaw2020bve}:

\begin{equation}
\label{eq:pderivative}
    PDQ_{do,di}(a, b) = \frac{d_o(P(a), P(b))}{d_i(a, b)}, b \neq a
\end{equation}

Note that high derivative values indicate a pair of very dissimilar output (high numerator) with similar inputs (low denominator), hence revealing sets of input values that are more sensitive to changes in the software behaviour. Going back to our Date example, let us consider the $d_i$ and $d_o$ for Dates as the edit distance~\footnote{The edit distance between two strings $A$ and $B$ is the number of operations (add, remove or replace a character) required to turn string $A$ into the string $B$.} between the inputs and outputs, respectively, when seen as strings (note that valid dates are just printed back as strings on the output side): 

\begin{itemize}
    \item \texttt{i1 = "2021-03-31"; P(i1) = "2021-03-31"}.
    \item \texttt{i2 = "2021-04-31"; P(i2) = "Invalid date"}.
    \item \texttt{i3 = "2021-04-30"; P(i3) = "2021-04-30"}.
\end{itemize}

\begin{figure}
    \centering
    \includegraphics[width=0.8\linewidth]{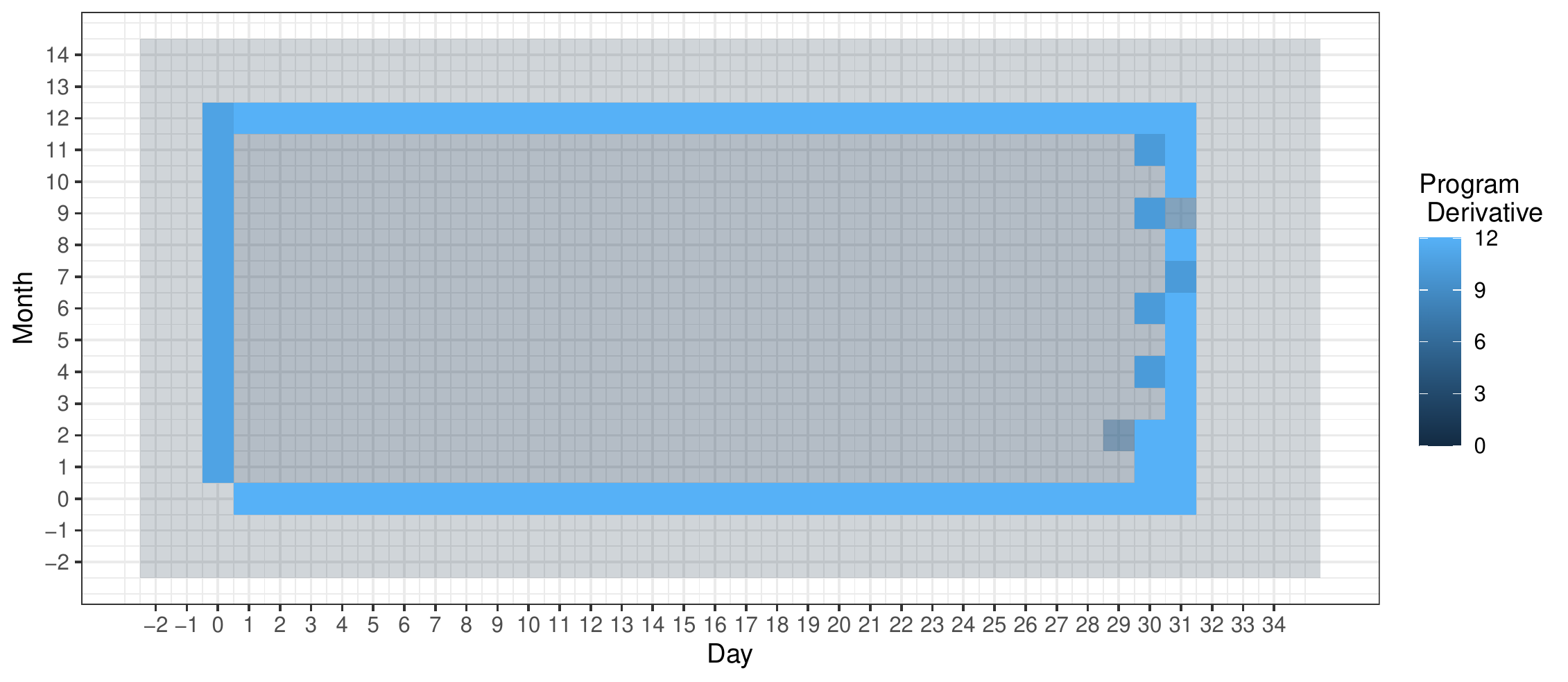}
    \caption{Plot of the program derivative by varying days and month for the year 2021. Higher opacity and brighter colors indicate higher derivative values. The derivative indicates the values that divide the boundary between valid and invalid combinations of days and months. Figure adapted from~ \cite{dobslaw2020bve}.}
    \label{fig:bve_example}
\end{figure}

As a consequence, $d_i(i1, i2) = 1$ as only one character changes between those input, whereas the output distance $d_o(P(i1), P(i2)) = 12$ since there is no overlap between the outputs, resulting in the $PD = 12/1 = 12$. In contrast, the derivative $PD(i1,i3) = 2/2 = 1$ is significantly lower and does not indicate any sudden changes in the output. In other words, the derivative changes significantly for i1 and i2, indicating boundary behavior. Figure~\ref{fig:bve_example} illustrates the program derivative of our example by varying months and dates for a fixed year value (2021) for a typical Date library. We see that automated boundary value testing can help highlight and, here, visualise boundary values. 

Note that the high program derivative values delimits the boundaries for the input on those two dimensions. Therefore, program derivative is a promising candidate to be a fitness function to identify boundary values in the input space. Using our BMI example, note that we can use the program derivative to identify the pairs of height and weight that trigger changes between classifications by comparing the variation in the output distance of similar input values. For instance, the output classifications can change, e.g., from "Overweight" to "Obese" by comparing individuals of same height but different weight values.

However, there are still many challenges when automating the generation of boundary values. First, software input is often complex and non-numerical such as objects or standardised files, which introduces the challenge of defining a suitable and accurate distance function able to measure the distances between input\slash output values. Second, the input space can have many dimensions (e.g., several input arguments) of varied types and constraints such that searching through that space is costly and sometimes infeasible. Last, but not least, boundary values often involve the tester's domain knowledge or personal experience that are hard to abstract in terms of functions or quantities (e.g., think of the Millennium bug for the date 2000-01-01). More important than fully automating the search of boundaries, testers are encouraged to employ boundary value exploration (BVE) techniques. BVE is a set of techniques (e.g., the visualisation in Figure~\ref{fig:bve_example}) that propose sets of candidate boundary values to help testers in refining their knowledge of boundaries in their own programs under test\cite{dobslaw2020bve}.

\subsection{Finding Diverse Test Suites}


A natural intuition that we have as software testers is that the tests we write and run need to differ from each other for the system to be properly tested. If we repeatedly rerun the same test case, or some set of very similar test cases, they are unlikely to uncover unique behaviours of the tested system. All of these similar tests will tend to pass or fail at the same time. Many AI-based techniques---including search-based approaches---have been proposed to select a good and complementary set of test cases, i.e. a diverse test suite. For example, recent research uses reinforcement learning to adapt the strategy employed by a search-based algorithm to generate more diverse tests for particular classes-under-test~\cite{Almulla21:AFFS}. A study comparing many different test techniques found that techniques focused on diversity were among the best possible in selecting a small set of test cases~\cite{henard2016comparing}.

A key problem in applying AI to find diverse test suites is how to quantify diversity. There are many ways in which we can measure how different test cases are, i.e. such as their length, which methods of the tested system they call, which inputs they provide etc. A general solution is to use general metrics from the area of Information Theory that can be used regardless of the type of data, length or specifics of the test cases we want to analyse. One study showed how metrics based on compression were very useful in quantifying test case diversity~\cite{feldt2016test}. Their experiments also showed that test sets comprised of more diverse test cases had better coverage and found more faults. 

A potential downside of these methods are that they can be expensive in terms of computations; many test cases and sets need to be considered to find the most diverse ones. Later research have proposed ways to speed diversity calculations up. One study used locality-sensitive hashing to speed up the diversity calculations~\cite{miranda2018fast}. Another study used the pair-wise distance values of all test cases as input to a dimensionality reduction algorithm so that a two-dimensional (2D) visual ``map'' of industrial test suites could be provided to software engineers~\cite{neto2018visualizing}.

\subsection{Oracle Generation and Specification Mining}


This chapter has focused on automatically generating the test inputs and test actions of good test cases. This excludes a key element of any test case: how to judge if the behavior of the system under test is correct. Can AI techniques help us also in generating oracles that make these judges? Or more generally, can we find or extract, i.e. mine, a specification of the SUT from actual executions of it?

Oracle generation is notoriously difficult and likely cannot be solved once and for all. While attempts have been made to ``learn'' a full oracle using supervised learning techniques, they are typically only viable on small and simple code examples.\footnote{An overview of attempts to use machine learning to derive oracles is offered by Fontes and Gay:~\cite{Fontes21:Oracles}.} Still, some researchers have proposed that AI can at least partly help~\cite{langdon2017inferring}. For example, one study used the Deep AI technique of neural embeddings to summarise and cluster the execution traces of test cases~\cite{tsimpourlas2021embedding}. Their experiments showed that the embeddings were helpful in classifying test case executions as either passing and failing. While this cannot directly be used as an oracle it can be used to select test cases to show to a human tester which can then more easily judge if the behavior is correct or not. Such interactive use of optimisation and AI in software testing has previously been shown to be effective~\cite{marculescu2015initial}. 

\subsection{Other AI Techniques}


Many other AI and Machine Learning techniques beyond those that we have described in this chapter have been used to support unit testing tasks, from input generation, to test augmentation, to test selection during execution. The trend is also that the number of such applications grows strongly year by year. Below we provide a few additional examples.

Researchers have proposed the use of Reinforcement Learning when generating test inputs~\cite{huurman2020generating}. They implemented the same test data generation framework that had been previously used with traditional search-based, meta-heuristics~\cite{feldt2013finding} as well as with Nested Monte-Carlo Search~\cite{poulding2015automated} but instead used Reinforcement Learning to generate new test cases. A neural net was used to model the optimal choices when generating test inputs for testing a system through its API. Initial results showed that technique could reach higher coverage for larger APIs where more complex scenarios are needed for successful testing. Another early study showed how Deep Reinforcement Learning could develop its own search-based algorithm that achieves full branch coverage on a training function and that the trained neural network could then achieve high coverage also on unseen tested functions~\cite{kim2018generating}. This indicates that modern AI techniques can be used to learn transferable testing skills. 

Reinforcement learning has also been used \textit{within} search-based test generation frameworks to adapt the test generation strategy to particular systems or problems. For example, it has been applied to automatically tune parameters of the metaheuristic~\cite{Jia2015}, to select fitness functions in multi-objective search in service of optimising a high-level goal (e.g., selecting fitness functions that cause a class to throw more exceptions)~\cite{Almulla21:AFFS}, and to transform test cases by substituting individual actions for alternatives that may assist in testing inheritance in class hierarchies or covering private code~\cite{He15:RL}

Other researchers have proposed the use of supervised machine learning to generate test input (e.g.,~\cite{Budnik18:SystemML,Walkinshaw17:QueryStrategy}). In such approaches, a set of existing test input and results of executing that input (either the output or some other result, such as the code coverage) are used to train a model. Then, the model is used to guide the selection of new input that attains a particular outcome or interest (e.g., coverage of a particular code element or a new output). It has been suggested that such approaches could be useful for boundary identification---Budnik et al. propose an exploration phase where an adversarial approach is used to identify small changes to input that lead to large differences in output, indicating boundary areas in the input space where faults are more likely to emerge~\cite{Budnik18:SystemML}. They also suggest comparing the model prediction with the real outcome of executing the input, and using misclassifications to indicate the need to re-train the model. Such models may also be useful for increasing input diversity as well, as prediction uncertainty indicates parts of the input space that have only been weakly tested~\cite{Walkinshaw17:QueryStrategy}. 




%% file: conclusion.tex
Unit testing is a popular testing practice where the smallest segment of code that can be tested in isolation from the rest of the system---often a class---is tested. Unit tests are typically written as executable code, often in a format provided by a unit testing framework such as \texttt{pytest} for Python. 

Creating unit tests is a time and effort-intensive process with many repetitive, manual elements. Automation of elements of unit test creation can lead to cost savings and can complement manually-written test cases. To illustrate how AI can support unit testing, we introduced the concept of search-based unit test input generation. This technique frames the selection of test input as an optimization problem---\textit{we seek a set of test cases that meet some measurable goal of a tester}---and unleashes powerful metaheuristic search algorithms to identify the best possible test input within a restricted timeframe. 

Readers interested in the concepts explored in this chapter are recommended to read further on the advanced concepts, such as distance-based fitness functions, multi-objective optimization, generating human-readable input, finding input boundaries, increasing suite diversity, oracle generation, and the use of other AI techniques---such as machine learning---to generate test input.

%% file: main.bbl
\begin{thebibliography}{10}

\bibitem{Google20:AndroidTesting}
Android Developers.
\newblock Fundamentals of testing.
\newblock 2020.
\newblock \url{https://developer.android.com/training/testing/fundamentals}.

\bibitem{Luo14:EAF}
Qingzhou Luo, Farah Hariri, Lamyaa Eloussi, and Darko Marinov.
\newblock An empirical analysis of flaky tests.
\newblock In {\em Proceedings of the 22Nd ACM SIGSOFT International Symposium
  on Foundations of Software Engineering}, FSE 2014, pages 643--653, New York,
  NY, USA, 2014. ACM.

\bibitem{eck2019understanding}
Moritz Eck, Fabio Palomba, Marco Castelluccio, and Alberto Bacchelli.
\newblock Understanding flaky tests: The developer’s perspective.
\newblock In {\em Proceedings of the 2019 27th ACM Joint Meeting on European
  Software Engineering Conference and Symposium on the Foundations of Software
  Engineering}, pages 830--840, 2019.

\bibitem{McMinn04:SBTesting}
Phil McMinn.
\newblock Search-based software test data generation: A survey.
\newblock {\em Software Testing, Verification and Reliability}, 14:105--156,
  2004.

\bibitem{Lukascyzk20:Python}
Stephan Lukasczyk, Florian Kroi{\ss}, and Gordon Fraser.
\newblock Automated unit test generation for python.
\newblock In Aldeida Aleti and Annibale Panichella, editors, {\em Search-Based
  Software Engineering}, pages 9--24, Cham, 2020. Springer International
  Publishing.

\bibitem{Arcuri13:Normalize}
Andrea Arcuri.
\newblock It really does matter how you normalize the branch distance in
  search-based software testing.
\newblock {\em Software Testing, Verification and Reliability}, 23(2):119--147,
  2013.

\bibitem{mao2016sapienz}
Ke~Mao, Mark Harman, and Yue Jia.
\newblock Sapienz: Multi-objective automated testing for android applications.
\newblock In {\em Proceedings of the 25th International Symposium on Software
  Testing and Analysis}, pages 94--105, 2016.

\bibitem{hierons2002comparing}
R.M. Hierons.
\newblock Comparing test sets and criteria in the presence of test hypotheses
  and fault domains.
\newblock {\em ACM Transactions on Software Engineering and Methodology
  (TOSEM)}, 11(4):448, 2002.

\bibitem{poulding2015automated}
Simon Poulding and Robert Feldt.
\newblock The automated generation of humancomprehensible xml test sets.
\newblock In {\em Proc. 1st North American Search Based Software Engineering
  Symposium (NasBASE)}, 2015.

\bibitem{silver2016mastering}
David Silver, Aja Huang, Chris~J Maddison, Arthur Guez, Laurent Sifre, George
  Van Den~Driessche, Julian Schrittwieser, Ioannis Antonoglou, Veda
  Panneershelvam, Marc Lanctot, et~al.
\newblock Mastering the game of go with deep neural networks and tree search.
\newblock {\em nature}, 529(7587):484--489, 2016.

\bibitem{McMinn10:humanoracle}
Phil McMinn, Mark Stevenson, and Mark Harman.
\newblock Reducing qualitative human oracle costs associated with automatically
  generated test data.
\newblock In {\em Proceedings of the First International Workshop on Software
  Test Output Validation}, STOV '10, pages 1--4, New York, NY, USA, 2010. ACM.

\bibitem{Alsharif2019:HOC-SQL}
Abdullah Alsharif, Gregory~M. Kapfhammer, and Phil McMinn.
\newblock What factors make sql test cases understandable for testers? a human
  study of automatic test data generation techniques.
\newblock In {\em International Conference on Software Maintenance and
  Evolution (ICSME 2019)}, pages 437--448, 2019.

\bibitem{chen2021evaluating}
Mark Chen, Jerry Tworek, Heewoo Jun, Qiming Yuan, Henrique Ponde, Jared Kaplan,
  Harri Edwards, Yura Burda, Nicholas Joseph, Greg Brockman, et~al.
\newblock Evaluating large language models trained on code.
\newblock {\em arXiv preprint arXiv:2107.03374}, 2021.

\bibitem{pearce2021empirical}
Hammond Pearce, Baleegh Ahmad, Benjamin Tan, Brendan Dolan-Gavitt, and Ramesh
  Karri.
\newblock An empirical cybersecurity evaluation of github copilot's code
  contributions.
\newblock {\em arXiv preprint arXiv:2108.09293}, 2021.

\bibitem{felt2020derivative}
Robert Feldt and Felix Dobslaw.
\newblock Towards automated boundary value testing with program derivatives and
  search.
\newblock In Shiva Nejati and Gregory Gay, editors, {\em Search-Based Software
  Engineering}, pages 155--163, Cham, 2019. Springer International Publishing.

\bibitem{dobslaw2020bve}
Felix Dobslaw, Francisco~Gomes de~Oliveira~Neto, and Robert Feldt.
\newblock Boundary value exploration for software analysis.
\newblock In {\em 2020 IEEE International Conference on Software Testing,
  Verification and Validation Workshops (ICSTW)}, pages 346--353, 2020.

\bibitem{Almulla21:AFFS}
Hussein Almulla and Gregory Gay.
\newblock Learning how to search: Generating effective test cases through
  adaptive fitness function selection.
\newblock {\em CoRR}, abs/2102.04822, 2021.

\bibitem{henard2016comparing}
Christopher Henard, Mike Papadakis, Mark Harman, Yue Jia, and Yves Le~Traon.
\newblock Comparing white-box and black-box test prioritization.
\newblock In {\em 2016 IEEE/ACM 38th International Conference on Software
  Engineering (ICSE)}, pages 523--534. IEEE, 2016.

\bibitem{feldt2016test}
Robert Feldt, Simon Poulding, David Clark, and Shin Yoo.
\newblock Test set diameter: Quantifying the diversity of sets of test cases.
\newblock In {\em 2016 IEEE International Conference on Software Testing,
  Verification and Validation (ICST)}, pages 223--233. IEEE, 2016.

\bibitem{miranda2018fast}
Breno Miranda, Emilio Cruciani, Roberto Verdecchia, and Antonia Bertolino.
\newblock Fast approaches to scalable similarity-based test case
  prioritization.
\newblock In {\em 2018 IEEE/ACM 40th International Conference on Software
  Engineering (ICSE)}, pages 222--232. IEEE, 2018.

\bibitem{neto2018visualizing}
Francisco Gomes De~Oliveira Neto, Robert Feldt, Linda Erlenhov, and Jos{\'e}
  Benardi De~Souza Nunes.
\newblock Visualizing test diversity to support test optimisation.
\newblock In {\em 2018 25th Asia-Pacific Software Engineering Conference
  (APSEC)}, pages 149--158. IEEE, 2018.

\bibitem{Fontes21:Oracles}
Afonso Fontes and Gregory Gay.
\newblock Using machine learning to generate test oracles: A systematic
  literature review.
\newblock In {\em Proceedings of the 1st International Workshop on Test
  Oracles}, TORACLE 2021, page 1–10, New York, NY, USA, 2021. Association for
  Computing Machinery.

\bibitem{langdon2017inferring}
William~B Langdon, Shin Yoo, and Mark Harman.
\newblock Inferring automatic test oracles.
\newblock In {\em 2017 IEEE/ACM 10th International Workshop on Search-Based
  Software Testing (SBST)}, pages 5--6. IEEE, 2017.

\bibitem{tsimpourlas2021embedding}
Foivos Tsimpourlas, Gwenyth Rooijackers, Ajitha Rajan, and Miltiadis Allamanis.
\newblock Embedding and classifying test execution traces using neural
  networks.
\newblock {\em IET Software}, 2021.

\bibitem{marculescu2015initial}
Bogdan Marculescu, Robert Feldt, Richard Torkar, and Simon Poulding.
\newblock An initial industrial evaluation of interactive search-based testing
  for embedded software.
\newblock {\em Applied Soft Computing}, 29:26--39, 2015.

\bibitem{huurman2020generating}
Steyn Huurman, Xiaoying Bai, and Thomas Hirtz.
\newblock Generating api test data using deep reinforcement learning.
\newblock In {\em Proceedings of the IEEE/ACM 42nd International Conference on
  Software Engineering Workshops}, pages 541--544, 2020.

\bibitem{feldt2013finding}
Robert Feldt and Simon Poulding.
\newblock Finding test data with specific properties via metaheuristic search.
\newblock In {\em 2013 IEEE 24th International Symposium on Software
  Reliability Engineering (ISSRE)}, pages 350--359. IEEE, 2013.

\bibitem{kim2018generating}
Junhwi Kim, Minhyuk Kwon, and Shin Yoo.
\newblock Generating test input with deep reinforcement learning.
\newblock In {\em 2018 IEEE/ACM 11th International Workshop on Search-Based
  Software Testing (SBST)}, pages 51--58. IEEE, 2018.

\bibitem{Jia2015}
Yue Jia, Myra~B. Cohen, Mark Harman, and Justyna Petke.
\newblock Learning combinatorial interaction test generation strategies using
  hyperheuristic search.
\newblock In {\em Proceedings of the 37th International Conference on Software
  Engineering - Volume 1}, ICSE '15, page 540–550. IEEE Press, 2015.

\bibitem{He15:RL}
W.~{He}, R.~{Zhao}, and Q.~{Zhu}.
\newblock Integrating evolutionary testing with reinforcement learning for
  automated test generation of object-oriented software.
\newblock {\em Chinese Journal of Electronics}, 24(1):38--45, 2015.

\bibitem{Budnik18:SystemML}
Christof Budnik, Marco Gario, Georgi Markov, and Zhu Wang.
\newblock Guided test case generation through ai enabled output space
  exploration.
\newblock In {\em Proceedings of the 13th International Workshop on Automation
  of Software Test}, AST '18, page 53–56, New York, NY, USA, 2018.
  Association for Computing Machinery.

\bibitem{Walkinshaw17:QueryStrategy}
N.~{Walkinshaw} and G.~{Fraser}.
\newblock Uncertainty-driven black-box test data generation.
\newblock In {\em 2017 IEEE International Conference on Software Testing,
  Verification and Validation (ICST)}, pages 253--263, March 2017.

\end{thebibliography}
